\documentclass[11pt]{article}

\usepackage{amsmath,amssymb}
\usepackage{bm}
\usepackage{wrapfig}
\usepackage[dvipdfmx]{graphicx}

\newcommand{\psibar}{\mbox{${\bar \psi}$}}

\newcommand\vp{{\bm{p}}}

\newcommand\eL{{\cal L}}

\newcommand\cA{{\cal A}}

\newcommand{\psla}{\mbox{\ooalign{\hfil/\hfil\crcr$p$}}}

\newcommand{\delsla}{\mbox{\ooalign{\hfil/\hfil\crcr{$\partial$}}}}

\newcommand{\rhChN}{\rho_c^{\rm NQ}}
\newcommand{\lk}{\left(}
\newcommand{\rk}{\right)}

\newcommand\beq{ \begin{eqnarray} }
\newcommand\eeq{ \end{eqnarray} }

\newcommand{\tpsp}{\hspace{1em}}

\title{\bf Axial-Vector and Tensor Spin Polarization  and Chiral Restoration in Quark Matter}

\author{
Tomoyuki Maruyama\\
College of Bioresource Sciences, Nihon University, Fujisawa 252-8510, Japan \\
\texttt{maruyama.tomoyuki@nihon-u.ac.jp} \\
Toshitaka Tatsumi
\\ Kitashirakawa Kamiikeda-cho 52-4, Kyoto 606-8287, Japan 
} 

\date{}

\begin{document}
\maketitle

\begin{abstract}
We study spontaneous spin-polarizations of quark matter with  flavour $SU(2)$ symmetry 
at zero temperature in the NJL model.  
In a relativistic framework, there are two types of the  spin-spin interactions: axial-vector (AV) and tensor (T), 
which accordingly give rise to different types of spin-polarized materials.
When the spin-spin interaction is sufficiently strong, 
the spin-polarized phase emerges within a specific density region.
As the spin-spin interaction becomes stronger, this phase extends over higher density region 
beyond the critical density of chiral restoration in normal quark matter.
We show that the spin-polarized phase leads to another kind 
of the spontaneous chiral symmetry breaking phase.
\end{abstract}


\section{Introduction}

The discovery of magnetars, neutron stars with extremely strong magnetic fields of the order of 
${\mathcal O}\lk 10^{15}\rk$~G, has reignited the long-standing question regarding the origin 
of intense magnetic fields in neutron stars  \cite{mag3,mag,mag2,ThDr95,MagRev1,MagRev2}. 
Various theories have proposed mechanisms for the generation and persistence of these magnetic fields throughout neutron star evolution \cite{MagRev1,MagRev2}, but a comprehensive explanation has yet to be established.

It was shown in studies  in Ref.~\cite{Braithwaite1,makishima14}  that 
 magnetars may contain both poloidal and toroidal magnetic field components within their interiors
with the latter being approximately $10^2$ times stronger than the former.
Then, it should be interesting to study the internal properties of hadronic matter as a microscopic origin of such strong magnetic fields in neutron stars.  
It may also provide some clue to solve the "magnetar" problem 

In recent years, theoretical and experimental research has extensively examined 
the QCD phase diagram in the temperature-density plane.
This diagram is believed to be intimately linked to phenomena observed in relativistic heavy-ion collisions
\cite{R-ion, SpPinHI, VMesonHI, SpinHI}, compact stars \cite{R-star1a, R-star1b, R-star1c}, 
and the early universe \cite{R-star2a, R-star2b}. 
 Notably, extensive studies have been  conducted on quark-gluon plasma (QGP)  
 in the high-temperature, low-density regime, 
and restoration of spontaneous breaking of chiral symmetry or color superconductivity 
in the high-density, low-temperature regime \cite{R-ion,SpinHI,csc}.

In dense matter, ferromagnetic properties can be realized when spin polarization (SP) occurs collectively 
among charged particles like baryons or quarks, highlighting the significance of spin interactions. 
The spontaneous spin polarization or magnetization of strongly interacting matter is important 
for understanding the origin of strong magnetic fields. In Ref.~\cite{tat00}, 
one of the authors (T.T.) proposed the possibility of a ferromagnetic transition in QCD. 
This transition could appear in quark matter due to one-gluon exchange interactions, with a critical density similar to nuclear density.

Using this idea,  we can estimate the magnetic field strength at the surface of compact stars. 
For a star with a mass of $M\sim 1.4 M_{\odot}$ and a radius of $R \sim 10$ km, 
and assuming a dipole field structure, the maximum surface magnetic field strength 
can  be estimated as $B_{\rm max}=(8\pi/3) f_Q\mu_q \rho_0$, 
where $f_Q$ epresents the quark matter volume fraction and $\mu_q$ the quark magnetic moment. 
In the extreme case, where $f_Q=1$, this gives $B_{\rm max} \sim O(10^{15-17})$ G, which agrees with observational data. 
This approach uses a perturbative calculation based on the Bloch mechanism, similar to methods in electron gas systems \cite{blo29,her66,Yoshida98,mor85}. 

Within the relativistic framework, we examine two types of spin density operators \cite{MaruTatsu00}: 
the first is the spatial component of the axial-vector (AV) current operator, defined as 
$\psi^\dagger \Sigma_i \psi \,(\equiv -\psibar \gamma_5 \gamma_{i}\psi)$, 
and the second is the spatial component of the tensor (T) operator, 
$\psi^\dagger \gamma^0 \Sigma_i \psi \,(\equiv -\frac{\epsilon_{ijk}}{2} \psibar \sigma_{jk} \psi)$, 
where $\psi$ represents the Dirac field. 
While these two operators are equivalent in the non-relativistic limit, they diverge significantly in behavior in the ultra-relativistic (massless) limit \cite{MaruTatsu00}. 
Throughout this text, we refer to the former as the AV-type spin operator and the latter as the T-type operator.

We analyzed the magnetic properties of quark matter using a non-perturbative method with AV-type and T-type spin operators. Starting with the AV interaction, we explored the spin-polarization (SP) mechanism under the mean-field approximation to show the possible coexistence of SP and color superconductivity (CSC) \cite{nak03A}, as well as the magnetic properties of an inhomogeneous chiral phase called the dual chiral density wave (DCDW) \cite{NT05}. Additionally, Maedan studied SP in the NJL model \cite{Namb04} using this AV mean field \cite{Maedan07}. In these systems, when the quark mass is kept constant, SP generally appears at high densities. However, the AV mean field weakens as the quark mass nears zero, suggesting that spin-polarized phases can only form within a limited density range.

Later, we have introduced the T interaction into quark matter and found that it behaves differently 
from the case of the AV interaction in manifesting spin-polarized phases. 
Unlike the AV-type SP, 
the T-type SP can appear even when the quark mass is zero \cite{MNT,MaruTatsu17}.  
In fact, Tsue et al. \cite{Tsue:2012nx} showed that SP arises in the density region 
beyond the critical density of spontaneous breaking of chiral symmetry of normal quark matter,
where the quark dynamical mass vanishes, using the NJL model with an effective potential approach.

Subsequently, we introduced the T interaction into quark matter and discovered distinct behavior in the emergence of spin-polarized phases compared to the AV interaction. 
Unlike AV-type SP, T-type SP  can develop even in the absence of quark mass \cite{MNT,MaruTatsu17}. 
Tsue et al. \cite{Tsue:2012nx} demonstrated that SP emerges in regions beyond the critical density for spontaneous chiral symmetry breaking in normal quark matter. 
In this chirally restored phase, where the quark dynamical mass vanishes, SP is achieved 
through the NJL model with an effective potential approach.

In these analyses, we adopted the mean-field approximation, incorporating AV and T channel interactions and introducing the corresponding mean fields. 
AV-type interactions in quark matter, for instance, can be derived from the perturbative one-gluon-exchange (OGE) interaction via the Fierz transformation. 
Conversely, T-type interactions likely stem from non-perturbative QCD effects, similar to those observed in effective hadron-hadron interactions \cite{Ring1, Shury}. 
Therefore, effective QCD models should account for both types of interactions. 
To date, various studies have explored the relationship between SP and other phases at high baryon-number densities, including the coexistence of SP with color superconductivity \cite{nak03A, Tsue:2012nx, Tsue:2014tra, Matsuoka:2016ufr}, as well as spatially homogeneous \cite{Maedan07,TPPY12,Matsuoka:2016spx,MaruTatsu17,Morimoto:2018pzk} and inhomogeneous chiral condensations \cite{NT05,Yoshiike:2015tha}. 
These studies indicate that AV- and T-type mean fields respond differently to dynamical chiral symmetry breaking: the AV-type spin-polarized phase cannot emerge when the dynamical quark mass is zero (i.e., in the chirally restored phase), whereas the T-type spin-polarized phase can still manifest. 
Consequently, examining the relationship between AV- and T-type SPs and their role in chiral symmetry restoration is essential.

Furthermore, in Fermi degenerate systems, magnetic fields are primarily coupled to the term proportional 
to the tensor density, $\psi^\dagger \gamma^0 \Sigma^i \psi$.
This suggests that AV- and T-type spin-polarized phases may exhibit distinct magnetic properties.

Interestingly, we shall see the deformation of the Fermi surface, which also differs 
between these two types of SP: 
in the AV-type phase, the Fermi surface is prolate, while in the T-type phase, it is oblate. 
This deformation is significant and may have large effects on the system's properties such as thermal transport or conductivity.  

To systematically investigate these effects, this paper employs the NJL model, treating AV and T interactions separately to analyze spin-polarized phases.

The organization of the paper is as follows. 
In the next section, we present the framework used for our analysis. 
Section 3 provides the results of the numerical calculations and discusses the relationship between spin polarization and chiral restoration. 
In this section, we omit vacuum polarization effects on spin polarization due to the unresolved ambiguity in its treatment. 
Finally, Section 4 summarizes our findings and offers concluding remarks.


\section{Formalism}
%
\subsection{Lagrangian}

In this section, we present our approach to modeling spin-polarized systems. 
A preferred direction arises along the axis of spin polarization, which we set to point in the positive $z$-direction. 
This setup breaks spherical symmetry, while preserving axial symmetry around the $z$-axis.

As introduced, 
spin-polarization in quark matter arises from two types of spin-spin interactions in the axial-vector (AV) and tensor (T) channels. 
We describe spin-polarized systems using the following Lagrangian density, 
which preserves flavor $SU(2)$ chiral symmetry:
\begin{equation}
\eL = \eL_{K} + \eL_S + \eL_V + \eL_A + \eL_T
\end{equation}
with
\begin{eqnarray}
\eL_{K} &=& \psibar (i \delsla - m) \psi ,
\\ 
\eL_S &=& - \frac{G_s}{2} \left[ (\psibar \psi )^2 
- ( \psibar \gamma_5 \tau \psi )^2 \right] ,
\label{Schn}
\\ 
\eL_V &=& - \frac{G_V}{2} \left[ (\psibar \gamma_\mu \psi )(\psibar \gamma^\mu \psi ) \right]
- \frac{G_{V}^{\prime}}{2} \left[ (\psibar \gamma_\mu \tau_a \psi )(\psibar \gamma^\mu \tau_a \psi ) \right] ,
\label{Vecn}
\\ 
\eL_A &=& - \frac{G_A}{2}  (\psibar\gamma_5  \gamma_\mu \psi )(\psibar \gamma^\mu \gamma_5 \psi )
- \frac{G_A^{\prime} }{2}  (\psibar\gamma_5  \gamma_\mu \tau_a \psi )(\psibar \gamma^\mu \tau_a \gamma_5 \psi ) ,
\label{AVcn}
\\
\eL_T &=&
- \frac{G_T}{2} \left[ 
(\psibar \sigma_{\mu \nu} \psi ) (\psibar \sigma^{\mu \nu} \psi ) 
- (\psibar \tau_a \gamma_5 \sigma_{\mu \nu} \psi ) 
(\psibar \tau_a \sigma^{\mu \nu} \gamma_5 \psi ) \right] ,
\label{TNcn}
\end{eqnarray}
where $\psi$ is the quark field operator, $m$ is the quark current mass, and $\tau_a$ ($a=1,2,3$) are the Pauli matrices in isospin space. The constants $G_s$ and $G_T$ are the coupling constants for the scalar and T channels, respectively, while $G_V$, $G_A$, $G_V^\prime$, and $G_A^\prime$ are the coupling constants for the isoscalar vector, isoscalar AV,  
isovector vector and isovector AV channels, respectively.

Here, we discuss the relationship between chiral symmetry and each interaction channel. 
The scalar channel of the Lagrangian respects chiral symmetry; however, only the scalar interaction generates a mean field, 
leading to spontaneous chiral symmetry breaking. 
Note that each term in this channel does not preserve $SU(2)$ chiral symmetry. 
Similarly, the T channel in Eq.~(\ref{TNcn}) satisfies $SU(2)$ chiral symmetry, 
but each term individually does not preserve any chiral symmetry, 
while all terms in the vector channel in Eq.~(\ref{Vecn}) and the AV channel in Eq.~(\ref{AVcn}) satisfy $SU(2)$ chiral symmetry. 
This suggests that the T mean field contributes to spontaneous chiral symmetry breaking.


\subsection{Quark Propagator and Mean-Field Approximation}

In this work, we focus on isospin-symmetric matter, where the densities of 
$u$-quark and $d$-quarrk are equal, and discard isovector terms from the Lagrangian.
 In addition, the isovector spin-polarized phases, where the polarized spin-direction
is opposite for  $u$- and $d$-quarks,  are possible even in isospin symmetric matter. 
 These phases are determined by the values and signs of the couplings, $G_A$, $G_A^\prime$ and $G_T$,
 but the expressions of the iso-scalar and isovector SPs are the same after the simple transformation,
 which is discussed later.
 Then, we, in the following, assume the iso-scalar spin-polarized phases in the actual expressions  
 by omitting $G_A^\prime$

Assuming a spatially uniform spin polarization,
the Dirac spinor $u({\vp,s})$ of the quark 
with momentum $\vp = (\vp_T, p_z)$ is derived as the solution of the following equation, 
\begin{equation}
\left[ \psla - M_q - U_0 \gamma^0 - U_A \gamma_0 \Sigma_z  -U_T \Sigma_z \right] u(\vp,s) = 0 ,
\label{Dirac-UT}
\end{equation}
within the mean-field approximation, 
with $\Sigma_z= {\rm diag}(1,-1,1,-1)$,  where the dynamical mass and each mean- field render 
\begin{eqnarray}
M_q &=& - G_s \rho_s  =- G_s < \psibar \psi> ,
\label{ScEq}
\\
U_0 &=& G_V \rho_q = G_V < \psibar\gamma^0 \psi> ,
\label{VcEq}
\\
U_A &=& G_A \rho_A  =G_A < \psibar\gamma^0 \Sigma_z \psi>,
\label{AvEq}
\\
U_T &=&  G_T \rho_T  = G_T <\psibar \Sigma_z \psi> .
\label{TsEq}
\end{eqnarray}

The vector mean field, $U_0$, serves solely to shift the single-particle energy and has no effect on spin properties.
Without the vector mean field $U_0$, the quark chemical potential $\mu$ does not increase monotonically with density, resulting in a density range that becomes inaccessible. 
In normal quark matter, the spontaneous breaking of chiral symmetry is restored at the density 
corresponding to the maximum chemical potential, with this transition being of first order. 
As the vector coupling strength $G_V$ increases, the transition density shifts higher, 
and, when $G_V$ is sufficiently large, the phase transition becomes second-order.

Thus, we can control the feature of the phase transition by tuning the strength of the vector interaction.
However, it does not essentially change properties of matter, especially the magnetic properties; 
it has a only role to exclude some density region in the chiral symmetry breaking phase. 
We, hereafter, assume that the vector coupling $G_V$ is large, and that 
the chiral phase transition is of the second order.
For simplicity, in addition, we rewrite $p_0 - U_0$ to 
$p_0$ and eliminate $U_0$ in the following formulation.

To proceed we introduce the quark propagator.
The quark propagator $S(p)$ is given by  the following  equation:
\begin{equation} 
{\tilde K} S(p) = 1 .
\label{prop1}
\end{equation} 
with 
\begin{equation} 
{\hat K} \equiv
\psla - M_q - U_A \gamma_0 \Sigma_z - U_T \Sigma_z .
\label{prop2}
\end{equation} 

The single-particle energy $p_0$ is obtained as a solution to the following equation:
\begin{eqnarray}
\det [{\hat K}]
&=& (p_0^2 - E_p^2)^2 - 2 p_0^2 (U_A^2 + U_T^2) - 8 p_0 M_q U_A U_T
- 2 M_q^2 (U_A^2 + U_T^2)
\nonumber \\ && 
 + (U_A^2 - U_T^2)^2 - 2 (U_A^2 - U_T^2) (p_z^2 - p_T^2) ~ = ~ 0 ,
\label{detAT} 
\end{eqnarray}
where $E_p = \sqrt{\vp^2 + M_q^2}$.   
This equation should yield two positive-energy solutions, $e_{\alpha=1,2}(\vp)$, and two negative-energy solutions, $e_{\alpha=3,4}(\vp)$, corresponding to the spin degrees of freedom. 
However, when $U_A \neq 0$ and $U_T \neq 0$, it becomes challenging to solve Eq.~(\ref{detAT}) analytically to find $e_\alpha(\vp)$.

The quark propagator is divided into a vacuum part, $S_F$, 
and a density-dependent part, $S_D$, 
which vanishes at zero temperature and density, as shown by
\begin{equation}
S(p) = S_F (p) + S_D (p).
\end{equation}
By using this separation, the densities in Eq.~(\ref{TsEq}) are also devided 
into the vacuum parts and the density dependent parts as
\begin{eqnarray}
\rho_{\Gamma_\alpha}\equiv< \psibar\Gamma_\alpha \psi> &=&
 N_d \int \frac{d^4 p}{(2 \pi)^4} {\rm Tr} \left[ i \Gamma_\alpha S(p) \right] 
\nonumber \\ &=&
 N_d \int \frac{d^4 p}{(2 \pi)^4} {\rm Tr} \left[ i \Gamma_\alpha  S_F (p) \right] 
+ N_d \int \frac{d^4 p}{(2 \pi)^4} {\rm Tr} \left[ i \Gamma_\alpha  S_D (p)  \right]  \quad
\nonumber \\ &=&
\rho_{\Gamma_\alpha}(V)+\rho_{\Gamma_\alpha}(D)
\end{eqnarray}
where $N_d = N_f N_c = 6$ represents the degeneracy factor, with $N_f = 2$ for the flavor
$N_c = 3$ for color degrees of freedom..

In addition, the quark density, which dose not include the vacuum part, is written as
\begin{equation}
 \rho_q = \sum_{\alpha=1,2} \int \frac{d^3 p}{(2 \pi)^3} n[ e_\alpha(\vp)]  
\label{RhQ}
\end{equation}
with $n(E)=\Theta(\mu - E)$ being the Fermi -Dirac distribution function, 
where $\Theta (x)$ is the step function, and $\mu$ is the quark number chemical potential.
Accordingly, the baryon density can be expressed as $\rho_B = \rho_q / N_c$.

The Fermionic part of thermodynamic potential now renders
\begin{eqnarray}
\Omega &=&N_d \int \frac{d^3p}{(2\pi)^3} \left[
\sum_{\alpha=1,2} \left( e_\alpha(\vp) -\mu \right) \Theta(\mu - e_\alpha(\vp)) + \sum_{\alpha=3,4} e(\vp,s) \right]
\nonumber \\ 
&=&
\Omega_D+\Omega_{vac}
\label{omega}
\end{eqnarray}
The density dependent term $\Omega_D$ is finite, while the vacuum contribution $\Omega_{vac}$ is infinite and some regularization is needed. 
In order to isolate the relevant vacuum contribution we apply 
the proper time regularization (PTR) \cite{SchPTR51}, 
\begin{eqnarray}
\Omega_{vac}
&=&N_d\int\frac{d^3p}{(2\pi)^3}\sum_{\alpha=3,4}e(\vp,s)
\nonumber \\
&=& -i N_d \int  \frac{d^4 p}{(2 \pi)^4}\sum_{\alpha=3,4}
\ln\left[p_0^2-\left(e_\alpha(\vp)+i\epsilon\right)^2\right]
%
\nonumber \\
&=& i N_d  \int_{1/\Lambda^2}^{\infty} \frac{d \tau}{\tau}  
\int  \frac{d^4 p}{(2 \pi)^4} \sum_{\alpha=3,4} 
e^{\tau\left[p_0^2-e_\alpha(\vp)^2\right]}
\label{ThPot}
\end{eqnarray}
 with $\Lambda$ being the cut-off parameter.
The vacuum parts of the  scalar, AV and T densities are then given by 
\begin{equation}
\rho_s (V)  = \frac{ \partial \Omega_{vac} }{ \partial M_q} , ~~
\rho_A (V)  = \frac{ \partial \Omega_{vac} }{ \partial U_A} , ~~
\rho_T (V)  = \frac{ \partial \Omega_{vac} }{ \partial U_T} .
\end{equation}

Before discussing the spin-polarized system, we should show the result 
for the spin-saturated (SS) system, $U_A = U_T = 0$.
In this system, the single particle energies can be written as $\pm e(\vp,s)=\pm\sqrt{\vp^2 + M_q^2}$ for each spin state $s=\pm 1$, and 
$\Omega_{vac}$ is written as
\begin{equation}
\Omega_{vac} = 2 N_d \int \frac{d^4 p_E}{(2 \pi)^4} \int_{1/\Lambda^2}^{\infty} 
\frac{d \tau}{\tau} e^{ - \tau (p_E^2 + M_q^2) }
=  \frac{N_d}{8 \pi^2}  \int_{1/\Lambda^2}^{\infty} \frac{d \tau}{\tau^3} e^{- \tau M_q^2 } ,
\end{equation}
where $p_E$ is Euclidian four dimensional momentum.
Then, the vacuum contribution of the scalar density is expressed as
\begin{eqnarray}
\rho_s (V) 
&=& \frac{- N_d}{4 \pi^2}  M_q  \int_{1/\Lambda^2}^{\infty} \frac{d \tau}{\tau^2} e^{- \tau M_q^2 }
= \frac{- N_d}{4 \pi^2}  M_q^3 \int_{M_q^2/\Lambda^2}^{\infty} \frac{d \tau}{\tau^2} e^{- \tau  }
\nonumber \\ &=&
\frac{- N_d}{4 \pi^2}  M_q \Lambda^2 F_2 \left( \frac{M_q^2}{\Lambda^2} \right)  .
\end{eqnarray}
where the function $F_\alpha$ is defined as
\begin{equation}
 F_\alpha (x) = x \int_{x}^{\infty} \frac{d \tau}{ \tau^\alpha } e^{- \tau } .
\label{Ffnc}
\end{equation}

On the other hand, the density dependent part of the scalar density is
\begin{eqnarray}
 \rho_s (D) 
&=& \frac{2 N_d }{(2 \pi)^3} \int_0^{p_F} d^3 p ~n \left[ \sqrt{p^2 + M_q^2} \right] \frac{M_q}{\sqrt{(p^2 + M_q^2)^3}}
\nonumber \\ &=& \frac{N_d}{(2 \pi)^2} M_q \left[ 2 p_F \mu - M_q^2 \ln \left(\frac{p_F + \mu}{M_q} \right) \right] ,
%
\end{eqnarray}
where
 $p_F = \sqrt{\mu^2 - M_q^2}$ is the Fermi momentum in the SS system.

The dynamical quark mass is  given by 
\begin{equation}
1 + \frac{G_s (\rho_s(V)+\rho_s(D))}{M_q}  =0 ,
\label{NJLEqS}
\end{equation}
Then, the dynamical quark mass in the vacuum, $M_0$, is given by
\begin{equation}
1 - \frac{ N_d}{4 \pi^2}  G_s  \Lambda^2 F_2 \left( \frac{M_0^2}{\Lambda^2} \right)  =0 ,
\label{dyn}
\end{equation}
and the critical density of the chiral restoration, $\rhChN$, is obtained as
\begin{equation}
1 - \frac{ N_d}{4 \pi^2} G_s \left( \Lambda^2 - 2 p_c^2 \right)  =0 
\end{equation}
with $\rhChN =  N_d p_c^3 /3 \pi^2$ and $F_2 (0) = 1$, 
which is the fiducial density in the following discussion. 
Note that there is a trivial solution $M_0=0$ besides the one in Eq.~(\ref{dyn}), 
while the latter solution is energetically favored in normal quark matter.

\bigskip
Next, we explain the formulation in the spin-polarized system.
The dynamical quark mass $M_q$ is also calculated with  Eq.~(\ref{NJLEqS}),
where the $U_A$ and $U_T$ are determined by
\begin{eqnarray}
1 - \frac{G_A \rho_A}{U_A} &=&0 ,
\label{NJLEqA}
\\
1 - \frac{G_T \rho_T}{U_T} &=&0 .
\label{NJLEqT}
\end{eqnarray}
However, the coexistence of the AV and T interactions  makes calculation
quite complicated.
In addition we do not have a sufficient information about the quark
interaction to distinguish the AV and T interactions, and there is no  serious reason to consider the both interactions  simultaneously. 
Thus we, hereafter, discuss the spin-polarization mechanism 
through the AV interaction and T interaction separately.


\subsection{AV Type Spin-Polarization}

\tpsp
From Eq.~(\ref{NJLEqA}), we can know that $\rho_A < 0$ when $G_A <0$ and $G_T = 0$.
In this subsection, then,  we consider this AV-type spin-polarized phases,
which we call  AVS.

The single particle energies $\pm  e(\vp,s)$ are expressed as
\begin{equation}
\pm e(\vp,s) = \pm\sqrt{(\sqrt{M_q^2+p_z^2} + s U_A )^2 + \vp_T^2}
 =\pm \sqrt{ E_p^2 + 2 s U_A \sqrt{ p_z^2 + M_q^2} + U_A^2 } .
\label{enSDA}
\end{equation}
where $E_p = \sqrt{\vp^2 + M_q^2}$ and
$s =\pm 1$ denotes the spin state of a quark,
By using the above single particle energies, the propagator of the quark is written as:
\begin{equation}
S(p) = S_F (p) + S_D (p),
\end{equation}
{where
\begin{eqnarray}
S_F(p) &=&\frac{\left[ \psla + M_q + \gamma^0 \Sigma_z U_A \right] 
\left\{ p^{2} - M_q^2 + U_A^2 
+ 2 U_A \left( p_0 \Sigma_z  - i p_x \alpha_y + i p_y \alpha_x \right) \right\} }
{(p_0^{2} - e^2(\vp,+1)+ i \delta)(p_0^{2} - e^2(\vp,-1)+ i \delta) }, \qquad
\label{SFAV}
\\
S_D(p) &=& \sum_{s=\pm 1}
\left[ \psla + M_q + \gamma^0\Sigma_z U_A \right] 
\left\{1 + \frac{ s  ( p_0 \Sigma_z  - i p_x \alpha_y + i p_y \alpha_x ) + s U_A}
{ \sqrt{p_z^2 + M_q^2}}\right\} 
\nonumber \\ && \quad \quad  \times
 \frac{i \pi}{2 e(\vp,s)} n(\vp,s) \delta[p_0 - e(\vp,s)] .
\label{SDAB}
\end{eqnarray}

The density dependent part of the scalar and AV densities are desribed as
\begin{eqnarray}
 \rho_s (D) &=& N_d \sum_{s=\pm 1} \int \frac{d^3 p}{(2 \pi)^3} 
n [e(\vp,s)] \frac{M_q }{e(\vp,s)  }\left[ 1 + \frac{ s U_A   }{\sqrt{p_z^2 + M_q^2}  } \right] ,
\label{RhASD}
\nonumber \\
\rho_A (D) &=& N_d \sum_{s=\pm 1} \int \frac{d^3 p}{(2 \pi)^3} 
n[ e(\vp,s) ] \frac{ s \sqrt{p_z^2 + M_q^2}  + U_A }{e (\vp,s) } .
\label{RhAD}
\end{eqnarray}

In addition, the detailed deviations of other densities  are presented in Appendix \ref{SecDDAV}.

The thermodynamical potential (\ref{ThPot})  is written as 
\begin{eqnarray}
\Omega_{vac} & = &   i N_d \sum_{s = \pm 1} \int \frac{d^4 p}{(2 \pi)^4} \int_{1/\Lambda^2}^{\infty} 
\frac{d \tau}{\tau} e^{\tau \left[ p_0^2 - p_T^2 - ( \sqrt{p_z^2 + M_q^2} + s U_A)^2 \right] }
%
%
\nonumber \\  &=&
 \frac{1}{ 8 \pi^{5/2} } \sum_{s = \pm 1} \int_{1/\Lambda^2}^{\infty} \frac{d \tau}{\tau^{5/2}}  
\int_{0}^{\infty} d p_z e^{- \tau \left( \sqrt{p_z^2 + M_q^2} + sU_A \right)^2 } 
\end{eqnarray}

The vacuum part of the scalar and AV densities are written as
\begin{eqnarray}
\rho_s(V) &=& \frac{ \partial \Omega_{vac}}{\partial M_q }
%
\nonumber \\  &=&  
- \frac{N_d \Lambda }{ 4 \pi^{5/2} } M_q \sum_{s = \pm 1}  \int_{0}^{\infty} d p_z    
 \left( 1 + \frac{sU_A}{\sqrt{p_z^2 + M_q^2} } \right)
 \nonumber \\ && \qquad\qquad\qquad\qquad
 \times
 F_{3/2} \left[ \left| \frac{ \sqrt{p_z^2 + M_q^2} + sU_A }{\Lambda} \right| \right] ,
 \label{SRVA}
\\ 
\rho_A(V) &=& \frac{ \partial \Omega_{vac}}{\partial U_A}
%
\nonumber \\  &=&  
  \frac{- N_d \Lambda}{ 4 \pi^{5/2} }  \sum_{s = \pm 1}  s
 \int_{0}^{\infty} d p_z  \left( \sqrt{p_z^2 + M_q^2} + sU_A \right)
 \nonumber \\ && \qquad\qquad\qquad\qquad
 \times
 F_{3/2}  \left[ \left| \frac{ \sqrt{p_z^2 + M_q^2} + sU_A }{\Lambda} \right| \right] ,
 \label{ARVA}
\end{eqnarray}
where $F_\alpha (x)$ is defined in Eq.~(\ref{Ffnc}).

Here, we examine the AV density when $M_q = 0$,  
The density dependent part of the AV density in Eq.~(\ref{RhAD}) becomse 
\begin{eqnarray}
\rho_A (D) &=& \sum_{s=\pm 1} \int \frac{d^3 p}{(2 \pi)^3} 
n[ e(\vp,s) ] \frac{ s |p_z|  + U_A }{ \sqrt{(p_z +  U_A)^2 + p_T^2} } .
\nonumber \\
 &=&  N_d \int \frac{d^3 p}{(2 \pi)^3} 
\left\{ n \left[ \sqrt{ (p_z + U_A)^2 + p_T^2 } \right] \frac{  p_z + U_A }
{ \sqrt{(p_z +  U_A)^2 + p_T^2} } 
 \right. \nonumber \\&& \qquad\qquad \quad ~ \left.
+ n \left[ \sqrt{ (-p_z + U_A)^2 + p_T^2 } \right]  
\frac{  -p_z + U_A}{ \sqrt{(p_z - U_A)^2 + p_T^2}} 
  \right\}
\nonumber \\
 &=&  N_d \int \frac{d^3 p}{(2 \pi)^3} n \left( \sqrt{p_z^2 + p_T^2} \right) 
\left\{  \frac{  p_z}{\sqrt{p_z^2 + p_T^2}} + \frac{ (-p_z) }{\sqrt{p_z^2 + p_T^2}}  \right\} ~=~ 0 .
\end{eqnarray}
The vacuum contribution in Eq.~(\ref{ARVA}) becomes
\begin{eqnarray}
\rho_A(V)  &=&  
  \frac{- N_d \Lambda}{ 4 \pi^{5/2} } 
 \int_{0}^{\infty} d p_z  \left\{ 
 \left( p_z + U_A \right) F_{3/2}  \left[ \left| \frac{ p_z + U_A }{\Lambda} \right| \right] 
-  \left( p_z - U_A \right) F_{3/2}  \left[ \left| \frac{ p_z - U_A }{\Lambda} \right| \right] 
\right\}
%
\nonumber \\   &=&  
  \frac{ N_d \Lambda}{ 4 \pi^{5/2} } 
 \int_{-U_A}^{U_A} d p_z    p_z F_{3/2}  \left[ \left| \frac{ p_z  }{\Lambda} \right| \right] 
= 0 .
\end{eqnarray}
When $U_A > 0$ and $M_q = U_T = 0$, the single particle energy satisfy the relation that 
$e(\vp_T, - p_z, -s )  = e(\vp_T, p_z, s )$.
The Fermi  sea can be visualized as two spheres with centers at $(0, 0, \pm U_A)$. 
These spheres contribute oppositely, resulting in the AV density of zero, $\rho_A =0$, and
the SP disappears.
Thus, when $M_q=0$, the AV density is zero $\rho_A(D)=\rho_A(V) = 0$, 
meaning the spin-polarized phase never appears in in the zero-mass quark regime.

\subsection{Tensor Type Spin-Polarization}

From Eq.~(\ref{NJLEqT}), we can see that $\rho_T < 0$ when $G_T < 0$ and $G_A = 0$.
In this subsection, we examine the T-type spin-polarized phases through the T interaction with $G_T < 0$ and $G_A = 0$.

The single particle energies  $\pm e(\vp,s)$ are given by
\begin{equation} 
\pm e(\vp,s) = \pm\sqrt{(\sqrt{M_q^2+\vp_T^2} + s U_T )^2 + p_z^2}
 =\pm \sqrt{ E_p^2 + 2 s U_T \sqrt{M_q^2+\vp_T^2} + U_T^2 }
\label{enSDT}
\end{equation}
with $s =\pm 1$.

The quark propagator is then written as:
\begin{equation}
S(p) = S_F (p) + S_D (p), 
\end{equation}
where 
\begin{eqnarray}
 S_F (p) &=&  \frac{\left[ \psla + M_q + \Sigma_z U_T \right] 
\left\{ p^2 - M_q^2 + U_T^2 
+ 2 U_T  ( p_z \gamma_5 \gamma^0 -  p_0 \gamma_5 \gamma^3 ) \right\} }
{ \left[ p_0^2 -e^2(\vp,1) + i \delta \right]
\left[ p_0^2 -e^2(\vp,-1) + i \delta \right] } ,
\\
S_D(p) &=& \sum_{s=\pm 1}
\left[ \psla + M_q + \Sigma_z U_T \right] 
\left\{1 + \frac{ s  ( p_z \gamma_5 \gamma^0 -  p_0 \gamma_5 \gamma^3 ) + s U_T}
{ \sqrt{\vp_T^2 + M_q^2}}\right\} 
\nonumber \\ && \quad \quad  \times
 \frac{i \pi}{2 e(\vp,s)} n(\vp,s) \delta[p_0 - e(\vp,s)] .
\end{eqnarray}

The  density dependent parts of the scalar and T densities are written as 
\begin{eqnarray}
\rho_s (D) &=& N_d
\sum_{s=\pm 1} \int \frac{d^3 p}{(2 \pi)^3}  ~n(\vp,s)~
\frac{M_q}{e(\vp,s)} \left(1  + \frac{s U_T} {\sqrt{M_q^2+p_T^2}} \right) .
\label{RhTSD}
%
\\
\rho_T(D) &=&
N_d  \sum_{s=\pm 1} \int \frac{d^3 p}{(2\pi)^3} ~ n(\vp,s)~
\frac{ s \sqrt{\vp_T^2 + M_q^2} + U_T }{e(\vp,s)} .
\label{RHTD}
\end{eqnarray}
Because $\rho_T <0$ when $U_T >0$,  so that Eq.~(\ref{NJLEqT})
has a solution when $G_T <0$.

In addition, the detailed deviations of other densities  are written in Appendix \ref{SecDDTS}.

The vacuum contribution of the thermodynamic potential density is written as 
\begin{eqnarray}
\Omega_{vac} & = &   i N_d \sum_{s = \pm 1} \int \frac{d^4 p}{(2 \pi)^4} \int_{1/\Lambda^2}^{\infty} 
\frac{d \tau}{\tau} e^{\tau \left[ p_0^2 - p_z^2 - ( \sqrt{p_T^2 + M_q^2} + s U_T)^2.\right] }
%
\nonumber \\  &=&
 \frac{N_d}{8 \pi^2} \sum_{s = \pm 1} \int_{1/\Lambda^2}^{\infty} 
\frac{d \tau}{\tau^2} \int_{0}^{\infty} d p_T p_T 
e^{- \tau (\sqrt{p_T^2 + M_q^2} + s U_T)^2 } .
\end{eqnarray}
The vacuum contribution of the  scalar density is then given by 
\begin{eqnarray}
\rho_s (V)  &=&  \frac{ \partial \Omega_{vac} }{ \partial M_q~~} =
 - \frac{N_d}{4 \pi^2} M_q \sum_{s = \pm 1} \int_{1/\Lambda^2}^{\infty} 
\frac{d \tau}{\tau} \int_{0}^{\infty} d p_T p_T \left( 1 + \frac{s U_T}{ \sqrt{p_T^2 + M_q^2} }\right) 
 \nonumber \\ && \qquad\qquad\qquad\qquad\qquad\qquad
 \times
e^{- \tau (\sqrt{p_T^2 + M_q^2} + s U_T)^2 } .
%
%
\nonumber \\ &=&
 - \frac{N_d}{4 \pi^2} M_q \sum_{s = \pm 1} \int_{1/\Lambda^2}^{\infty} 
\frac{d \tau}{\tau} \int_{M_q + s U_T}^{\infty} d x_T x_T  e^{- \tau x_T^2 } 
\quad 
\nonumber \\ &=&
 - \frac{N_d M_q}{8 \pi^2} \sum_s \int_{1/\Lambda^2}^{\infty} 
\frac{d \tau}{\tau^2} e^{- \tau (M_q + s U_T)^2 } 
\nonumber \\ &=&
  - \frac{N_d M_q}{8 \pi^2} \Lambda^2 \sum_s
F_2 \left( \frac{(M_q + s U_T)^2}{\Lambda^2} \right) ,
\end{eqnarray}

Here, we give a comment on the SP when $M_q =0$,

Different from the AV case, for $M_q =0$, the T density is non-zero
and described as 
\begin{eqnarray}
\rho_T(D) &=&
N_d  \sum_{s=\pm 1} \int \frac{d p_z d p_T p_T}{(2\pi)^2}  n [\sqrt{(p_T + sU_T)^2 + p_z^2}]
\frac{ s p_T+ U_T }{\sqrt{(p_T + sU_T)^2 + p_z^2}} 
\nonumber \\  &=&
\frac{N_d}{(2\pi)^2}  \sum_{s=\pm 1} s \int d p_z  \int_{s U_T} d p_T n \left( \sqrt{p_T^2 + p_z^2} \right)  
\frac{ p_T (p_T - s U_T) }{\sqrt{p_T^2 + p_z^2}} 
\nonumber \\  &=&
- \frac{N_d}{ 2 \pi^2}  \int d p_z  \left\{ \int_{U_T} d p_T n \left( |\vp| \right)  \frac{ p_T }{ |\vp| } 
+ \int_{0}^{U_T} d p_T n \left( |\vp| \right)  \frac{ p_T^2  }{ |\vp|}
\right\}  \neq 0 .
%
%
\\ 
\rho_T (V) &=&  \frac{N_d}{4 \pi^2} \Lambda^2  \left\{
2 \int_{0}^{U_T} d E_T F_1 \left( \frac{E_T^2}{\Lambda^2} \right) 
+  U_T F_2\left[\frac{U_T^2}{\Lambda^2}\right]  \right\} \neq 0 .
\end{eqnarray}
Thus, we can see that the T density is non-zero even when $M_q = 0$.

As mentioned before,  $M_q = 0$ is a  trivial solution of Eq.~(\ref{ScEq})  besides the dynamical mass
which is a  solution of Eq.~(\ref{NJLEqS}).
In the T type SP, then, the two kinds of  spin-polarized states can exist;
one is a state with $U_T > 0$ and $M_q > 0$, and the other is that with $U_T > 0$ and $M_q = 0$.
We will call the former  state TS-I and the latter state TS-II.

As previously mentioned, the purpose of this paper is to explore the relationship between spin polarization (SP) and chiral symmetry. To clarify this relationship, we set the current quark mass to zero ($m=0$) throughout this analysis.

Chiral symmetry is preserved only when both scalar and pseudo-scalar interactions are present. However, it is the scalar interaction that explicitly breaks this symmetry, leading to spontaneous chiral symmetry breaking in the mean-field approximation.

Regarding the relationship between spin polarization and chiral symmetry, it is important to note that the axial vector (AV) interaction respects chiral symmetry and does not break it. However, the AV mean-field induces spin polarization, which results in spontaneous chiral symmetry breaking. Consequently, the AV mean-field can exist when the quark acquires a dynamical mass.

Moreover, the tensor interaction considered in this work maintain $SU(2)$ chiral symmetry. In isospin-symmetric matter, tensor mean fields induce both spontaneous chiral symmetry breaking and spin polarization, even when the quark mass is zero.

\bigskip
\subsection{Remarks}

Here, we provide remarks on the vacuum contributions.
In our model with the PTR , the vacuum contributions of the AV and T densities 
are strongly dependent on $\Lambda$.
Then, these  give too  large negative contributions to the spin polarization,
and  never allow the spin-polarized phase to be realized..
However, the  QCD calculation indicate  that the negative magnetic  susceptibility 
at zero temperature limit is zero \cite{LQCD14} or negative \cite{LQCD12}.
The substantial vacuum contributions in our approach may not align with these findings.

Moreover, the vacuum terms in the AV and T densities depend significantly on the chosen regularization scheme and the specific cut-off value \cite{MaruTatsu17}.
For instance, in AV-type SP, the PTR regularization decreases SP \cite{NT05}, whereas a momentum cut-off regularization increases it \cite{Maedan07}.
In T-type SP, the PTR diminishes SP, while an effective potential approach enhances it \cite{TPPY12}.

The cut-off parameter limits the momentum space in the calculation.  
The T and AV densities are given by the difference between the spin-up and spin-down contributions, 
and then are sensitive to momentum space restrictions, 
making the results highly dependend on the regularization scheme and the value of the cut-off parameter.

In the usual renormalization procedure we renormalize the vacuum
polarization by introducing the suitable counter terms and redefine the T and AV densities to fit their physical values.
In the standard treatment of the NJL model, which is an unrenormalizable model, 
we simply regularize the vacuum contributions to the scalar density by using a cut-off parameter.
The vacuum part of the scalar density is associated with the dynamical quark mass in the vacuum, but nothing to do with the spin properties.
In the present model, the vacuum contributions of the AV and T densities strongly 
diverge as $\Lambda \rightarrow \infty$ even for small asymmetry of
the spin states. 
Thus it may be rather difficult to control the scalar and the AV and T densities simultaneously:
we do not have any clear rule to regularize the AV and T densities in a systematic way. 
Thus, the cut-off dependence of the AV and T densities from the vacuum
contribution is less meaningful in the present approach.
In the NJL model it is not easy to apply a consistent method 
even for the qualitative discussions.
In the next section, then, we perform actual calculation without 
the vacuum contributions for the AV and T densities: $\rho_{A,T} \approx \rho_{A,T} (D)$.

In standard renormalization, we handle vacuum polarization by adding counterterms, adjusting T and AV densities to reflect their physical values.
In the NJL model, which is inherently unrenormalizable, we control the vacuum contributions to the scalar density using a cut-off parameter.
The vacuum term in the scalar density correlates with the vacuum dynamical quark mass, rather than spin features.
In our model, the vacuum contributions to AV and T densities diverge as $\Lambda \rightarrow \infty$, even under slight spin-state asymmetry.
Thus, it becomes challenging to manage both scalar and AV/T densities concurrently.
We lack a systematic rule for regularizing AV and T densities, making the cut-off dependency of vacuum contributions in these densities less meaningful in our approach.
Applying a consistent method in the NJL model, even qualitatively, remains difficult.

Therefore, in the following section, we calculate the AV and T densities without including vacuum contributions: $\rho_{A,T} \approx \rho_{A,T} (D)$.

\bigskip

Furthermore, we provide an additional comment on the isospin dependence of spin polarization (SP).  
For the T-type SP, this holds only when the spin polarization is isoscalar, 
i.e., when the average spins of the $u$- and $d$-quarks are aligned in the same direction. 
In the case of isovector SP, the T densities of the $u$- and \( d \)-quarks have opposite signs.

In symmetric matter, we define 
$\rho_T = \rho_T(u) - \rho_T(d) = 2\rho_T(u)$ and $U_T = U_T(u) - U_T(d) = 2U_T(u)$,
which allows us to rewrite Eq.~(\ref{TsEq}) as $U_T = - G_T \rho_T$,
matching the form of Eq.~(\ref{NJLEqT}) except for the sign on the right-hand side.
This shows that when $G_T > 0$, the isovector SP can arise, 
with its strength comparable to the case where $G_T < 0$.

Although our discussion focuses on the case $G_T < 0$, 
the same argument applies to the isovector SP system when $G_T > 0$. 
In that scenario, the system exhibits enhanced ferromagnetic properties 
due to the opposite charges of the u- and d-quarks.

For the AV-type SP, the isoscalar and isovector axial vector couplings, $G_A$ and $G_A^\prime$, 
are unrelated to chiral symmetry. 
The SP is isoscalar when $G_A < G_A^\prime$ and $G_A < 0$, and isovector 
when $G_A^\prime < G_A$ and $G_A^\prime < 0$. SP does not occur when both $G_A > 0$ and $G_A^\prime > 0$.

Following a similar reasoning as for the T-type, we can describe the isovector SP 
by replacing $G_A$ with $G_A^\prime$.  
Therefore, we omit the case where $G_A > 0$ and $G_A^\prime > 0$,  
and in this work, we assume $G_A < 0$.

\section{Results}
\label{SecRes}

In this section we present numerical results examining SP in relation with chiral restoration.
For this purpose, we adopt the chiral limit and utilize two  parameter-sets 
from Ref.~\cite{NT05}, whose details are presented in Table~{\ref{Param}}.
 where we also show the chiral phase transition density in normal quark matter  $\rhChN$
in the unit of the nuclear saturation density ($\rho_0 = 0.17$ fm$^{-3}$).

\begin{table}
\begin{center}  
\caption{Parameter sets}\label{Param}
\begin{tabular}{|c|c|c|c|c|}
\hline 
~\textbf{Parameter} ~& ~$G_s \Lambda^2$~ & ~$\Lambda$ (MeV)~ & ~$M_0 / M_N$~ & $~\rho_{c}^{\rm NQ} / N_c \rho_0$~ 
\\ \hline PM1 & 6.0 & 850 & 0.426 & 3.4085
\\ \hline
PM2 & 6.35 & 660.37 & 0.352 & 1.7614  
\\ \hline
\end{tabular}
\end{center}  
\end{table}


In the spin-polarized system, spherical symmetry is broken.
Indeed, the expression of the single particle energies Eq.~(\ref{enSDA}) do not hold  spherical symmetry,
and thereby it causes the deformation of the Fermi sea, $n(e(\vp,s)) = \Theta(\mu - e(\vp,s))$
in the spin-polarized system.
\begin{figure}[hb]
\begin{center}
{\includegraphics[scale=0.4,angle=270]{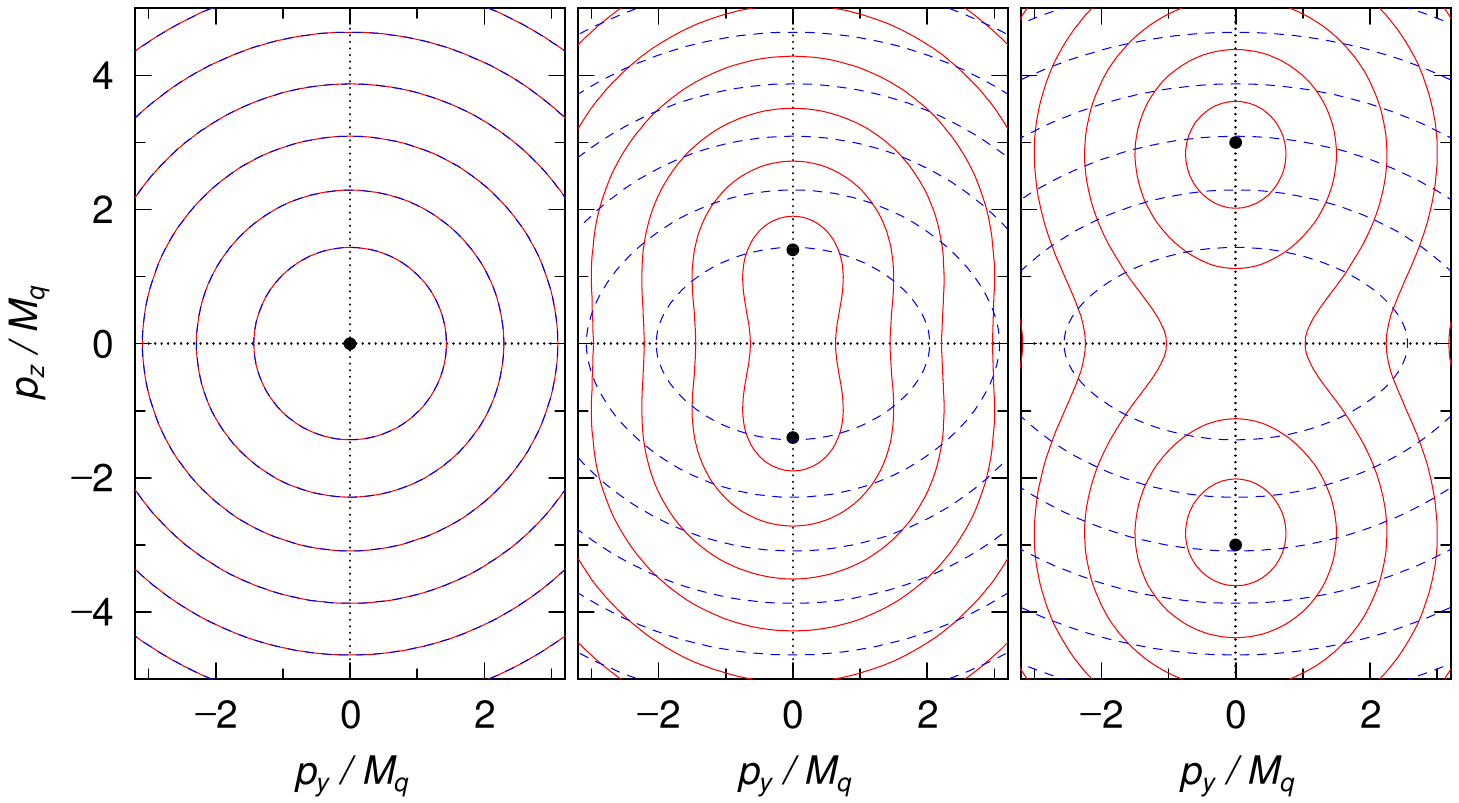}}
\noindent
\caption{\small Contour plots of the single particle energy minus its
  minimum value in the $p_y - p_z$ plane ($p_x = 0$)  when $U_A = 0$ (a), $U_A = 1.4 M_q$ (b) and $U_A = 2.8 M_q$ (c).
The red solid and blue dashed lines show the results for the major spin states
 ($s=-1$) and that for the minor spin states ($s=1$).}
\label{SengAX}
\end{center}
\end{figure}

In Figure~\ref{SengAX} we show the contour plots of the single particle
energy,  $e(\vp,s) - e_{min}(s)$, where $e_{min}(s)$ is the minimum
energy of $e(\vp,s)$ as functions of momenta along the symmetry axis, $p_z$, 
and perpendicular to that, $p_y$ ($p_x =0$), which are scaled with the dynamical mass 
$M_q$ when $U_A = 0$ (a), $U_A = 1.4 M_q$ (b) and  $U_A = 2.8 M_q$ (c). 
The solid and dashed lines represent the results of $s=-1$ and $s=1$, respectively.
We can see  that the momentum distribution is
deformed  prolately when $s=-1$ and oblately when $s=1$.  
When $U_A > M_q$, the minimum of the single particle energy in the major
spin state ($s=-1$) becomes zero, $e_{min}(s=-1)=0$ and moves from
$\vp= 0$ to that $\vp = \left( 0, 0, \pm \sqrt{U_A^2 - M_q^2} \right)$.
It should be interesting to see that these points are known as the Weyl points 
in the context of the Weyl semimetal in condensed matter physics \cite{AMV2018,TYK2018} ,

They may play a peculiar role in the electric or thermal transport in the AV-type SP phase. 

\begin{figure}[bht]
\vspace*{-1.em}
\centering
{\includegraphics[scale=0.5,angle=270]{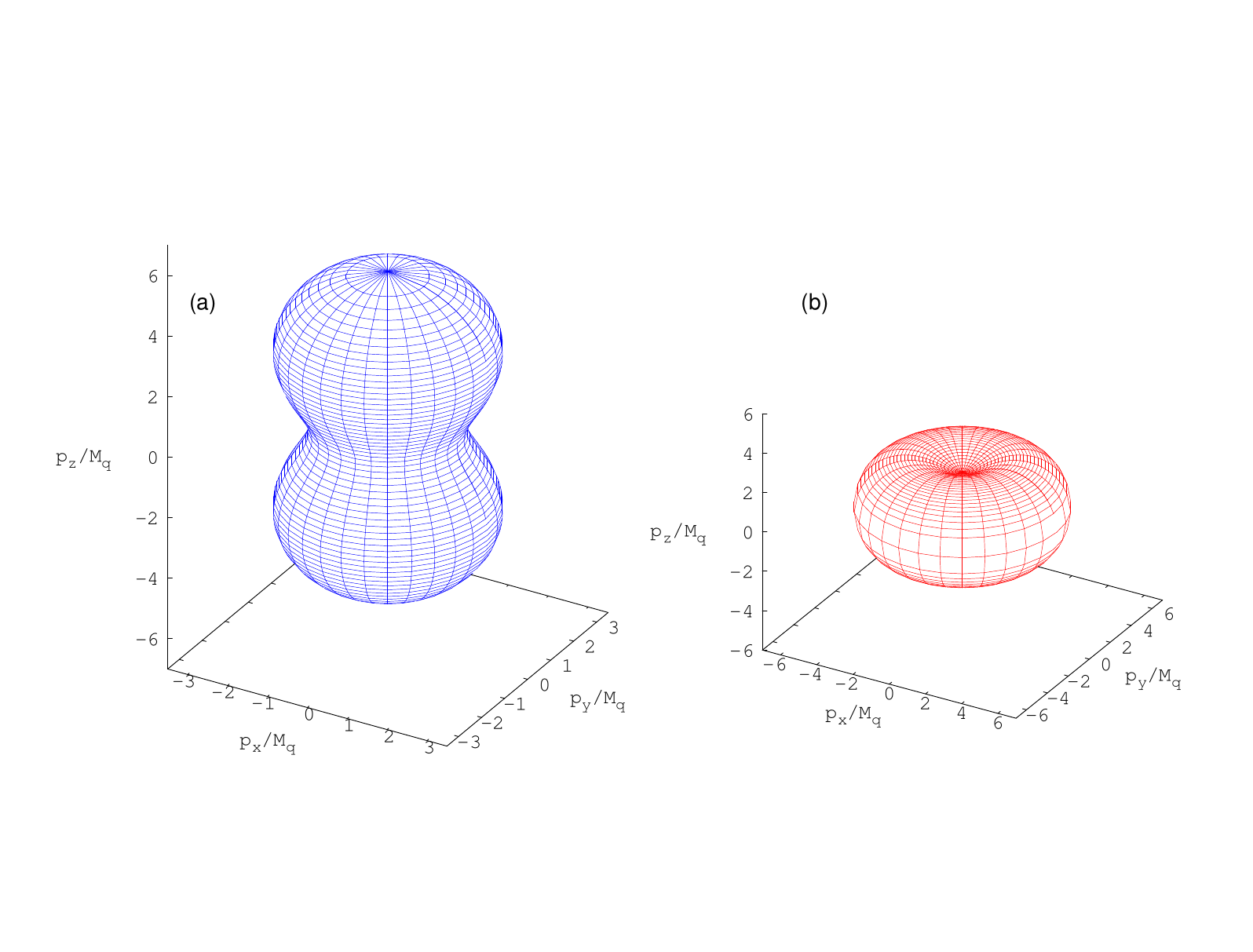}}
\vspace*{-4em}
\caption{The energy constant surfaces for $e = 2.5M_q$ and $s=-1$,
when $U_A = 2.8M_q$ (a) and when $U_T = 2.8M_q$ (b).}
\label{MomDis}
\end{figure}

In the T type spin-polarized system, 
the momentum distribution is deformed oblately when $s=-1$ and prolately 
when $s=1$   through the single particle energy, Eq.(\ref{enSDT}). 
Its shape is given by exchanging $p_y$ and
$p_z$ in Figure~\ref{SengAX}.
When $U_T>M_q$, $e_{min}(s=-1)$ becomes zero on the circle, $\vp=(p_x,p_y,0)$ with $p_x^2+p_y^2=U_T^2-M_q^2$, which corresponds to the line node in the context of topological material in condensed matter physics \cite{AMV2018,TYK2018} .

In order to clarify the difference in the Fermi distribution between the AV and T types,
we display the constant energy surface for $e(\vp,-1)  = 2.5 M_q$ in Figure~\ref{MomDis},
when $U_A=2.8M_q$ (a) or  $U_T=2.8M_q$ (b).
We see that difference in the momentum distribution between the two types of
the SP:
it is deformed prolately in the T-type SP and oblately in the AV-Type SP.

\bigskip

First, we discuss the AV-type SP.
\begin{figure}[tbh]
\centering
{\includegraphics[scale=0.43,angle=270]{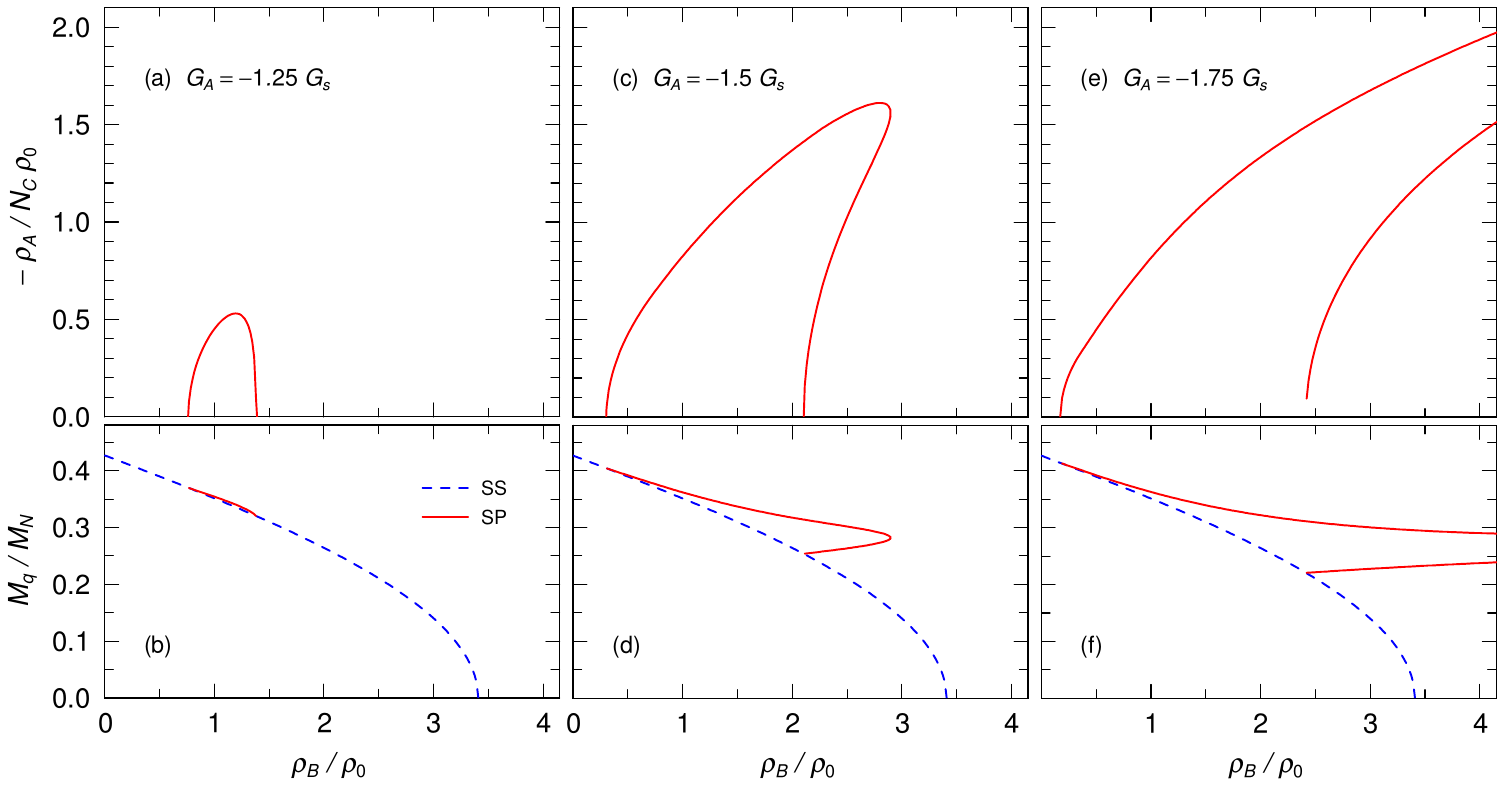}}
\caption{\small 
AV-type spin polarization characteristics with PM1.
In the upper panels (a, c, e), the tensor densities are normalized to the normal nuclear density.
In the lower panels (b, d, f), the dynamical quark mass is normalized by the nucleon mass 
for the spin-polarized (red solid lines) and spin-saturated (blue dashed line) phases.
Results for $G_A = -1.25$ (a, b), $=-1.5$ (c, d), and $=-1.75$ (e, f) are shown in the left, middle, and right panels, respectively.
}
\label{AVPM1}
\end{figure}

In Figure~\ref{AVPM1}, we show the AV density $\rho_A/(N_c \rho_0)$ with PM1  (upper panel) 
and that of the dynamical mass (lower panel)  versus baryon density  
for $G_A = - 1.25 G_s$ (a,b), $G_A = -1.5 G_s$ (c,d) and $G_A = -1.75 G_s$ (e,f),

The AV-type spin-polarized phase appears 
in $0.76 \lesssim \rho_B / \rho_0 \lesssim 1.38$ for $G_A = -1.25 G_s$ (a,b), 
and in $0.30 \lesssim \rho_B / \rho_0 \lesssim 2.89$ for $G_A = -1.5 G_s$ (c,d). 
For $G_A = -1.5 G_s$, there are two AVS at the same  density in  
$2.10 \lesssim \rho_B / \rho_0 \lesssim 2.89$. 
We cannot say which AVS is realized  without energetic discussions.
For $G_A = -1.75G_s$, the AVS  extends over a higher density region beyond
the chiral critical density, $\rho_B > \rhChN$.
As $-G_A$ increase, the AV density grows,  and  the spin-polarized phase
extends into higher density region beyond the chiral critical density in normal quark matter $\rho^{\rm NQ}_c$,  
where the dynamical quark mass is non-zero..

\begin{figure}[bht]
\centering
{\includegraphics[scale=0.43,angle=270]{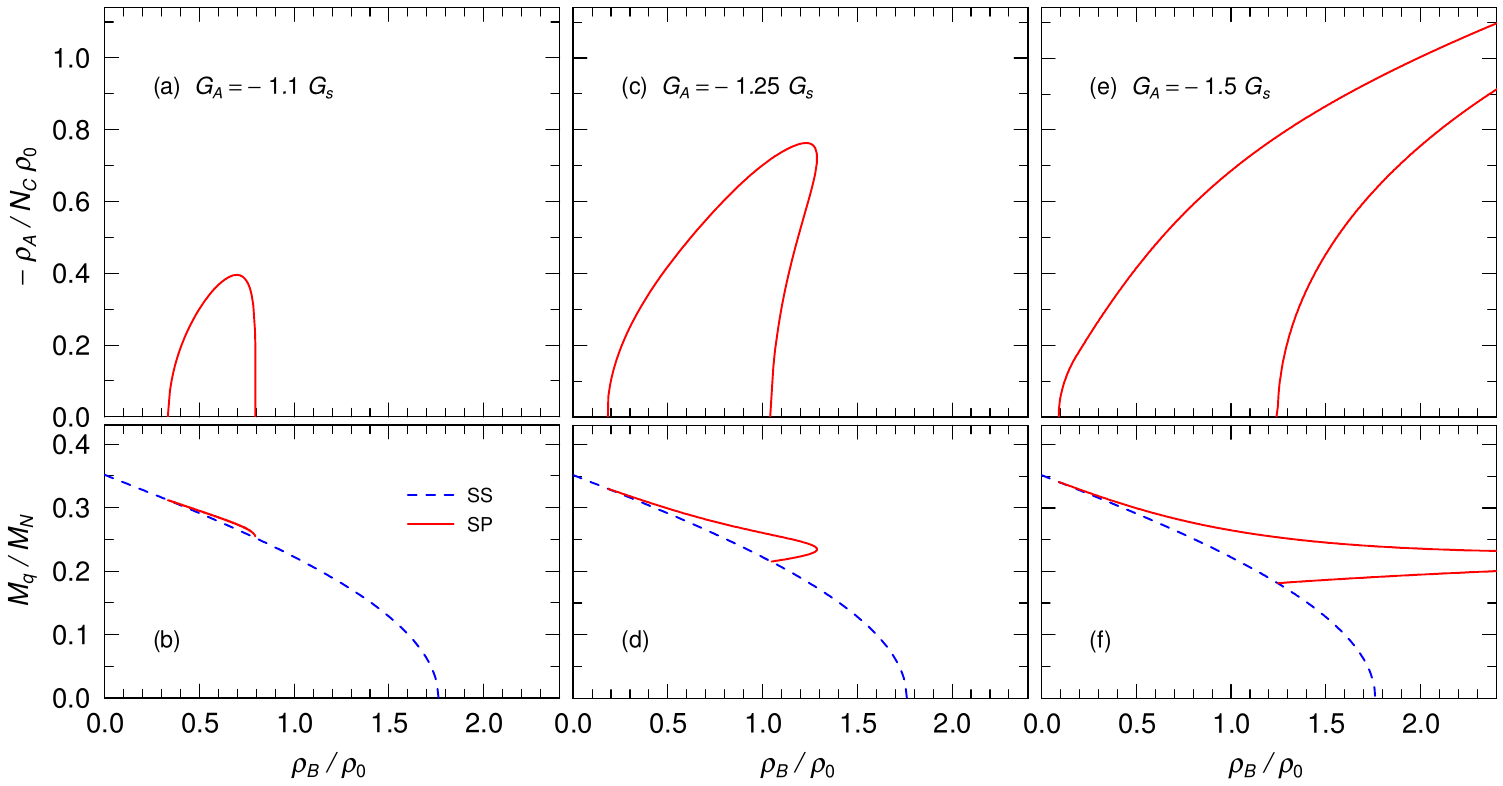}}
\caption{\small 
AV-type spin polarization characteristics with PM2.
In the upper panels (a, c, e), the AV densities are normalized to the normal nuclear density.
In the lower panels (b, d, f), the dynamical quark mass is normalized by the nucleon mass, shown 
for the spin-polarized (red solid lines) and spin-saturated (blue dashed line) phases.
Results for $G_A = -1.1$ (a, b), $=-1.25$ (c, d), and $=-1.75$ (e, f) are shown 
in the left, middle, and right panels, respectively.
}
\label{AVPM2}
\end{figure}

In Figure~\ref{AVPM2}, we show the results with the PM2 parameter set, where $G_s$ is about 1.06 times  
larger than that of PM1.
We note that the critical point of the coupling $G_A$, where the spontaneous SP appears, 
is between $-G_A/G_s =1.0$ and  $-G_A/G_s = 1.1$. 
The qualitative behaviors are very similar, but they appear in smaller $G_A$.

In Refs. \cite{NT05,Maedan07} the AV-type SP was mentioned to appear
only in narrow density region before the chiral restoration {($\rho_q < \rhChN$).
However, the present results show that the AV-type spin-polarized phases,
where the quark mass is non-zero,in the density region above $\rhChN$ 
when the AV interaction is large,

\bigskip
Next, we show the results about  the T-type SP.
In Figure~\ref{TSPM1}, we show the baryon density dependence 
of the tensor density $\rho_T/\rho_0$ with PM1 (upper panel) 
and that of the dynamical mass (lower panel) for
$G_T = - 1.25 G_s$ (a,b), $G_T = -1.5 G_s$ (c,d),  and $G_T = -1.75 G_s$ (e,f).
In the upper panel the solid lines represent $\rho_T/(N_c \rho_0)$ 
in the spin-polarized phase when $M_q > 0$ (TS-I), and the dotted lines indicate that 
when $M_q=0$ (TS-II). 
In the lower panel, the solid and dashed lines represent the dynamical quark mass 
in the spin-polarized and spin-saturated phases, respectively. 
Note again that some energetic discussions are necessary to see which phase is realized.

\begin{figure}[bht]
\begin{center}
{\includegraphics[scale=0.43,angle=270]{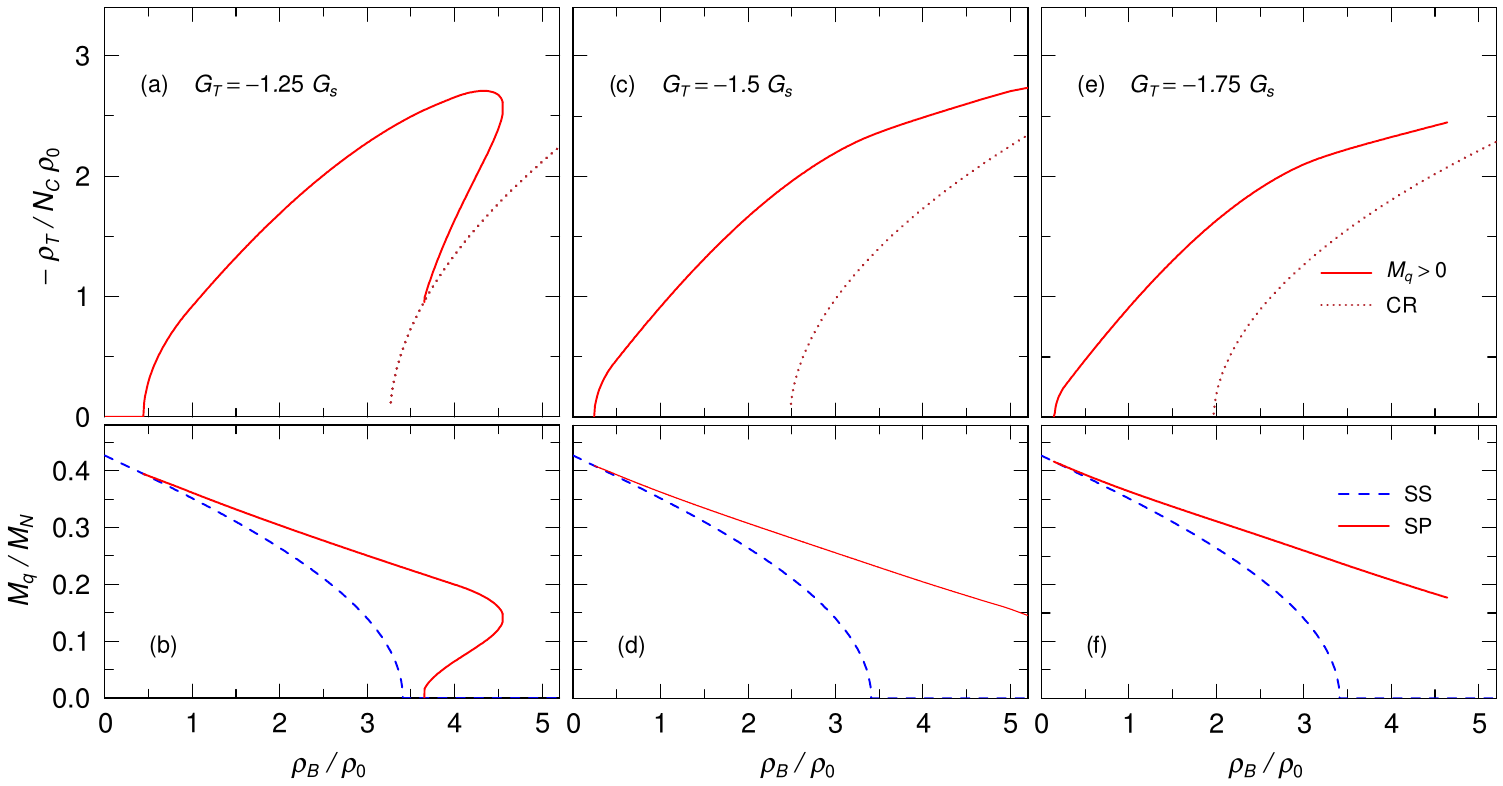}}
\caption{\small 
T-type spin polarization characteristics with PM1.
In the upper panels (a, c, e), the tensor densities are normalized to the normal nuclear density.
The solid and dotted lines correspond to the results in the massive quark and massless quark phases, respectively.
In the lower panels (b, d), the dynamical quark mass is normalized by the nucleon mass, shown 
for the spin-polarized (red solid lines) and spin-saturated (blue dashed line) phases.
The left, middle and right panels present the results for $G_T = -1.25 G_s$ (a, b), $=-1.5 G_s$ (c, d), and $=-1.75 G_s$ (e, f), respectively.
}
\label{TSPM1}
\end{center}
\end{figure}

For all the parameters, 
we can see that two kinds of spin-polarized phases, TS-I and TS-II appear,
and that the TS-I phase appears first, and 
 the TS-II phase appears in  the density region,  $\rho_B < \rhChN$.
For $G_T = -1.25 G_s$ there appear density regions where the three solutions 
corresponding to  the spin-polarized phases, 
two TS-I phases and one TS-II phase.
For $G_T = -1.5 G_s$ (c,d) and   $G_T = -1.75 G_s$, on the other hand, 
the TS-I phase  extends} to higher density region, and the end point is invisible, 

\begin{figure}[bht]
\centering
{\includegraphics[scale=0.43,angle=270]{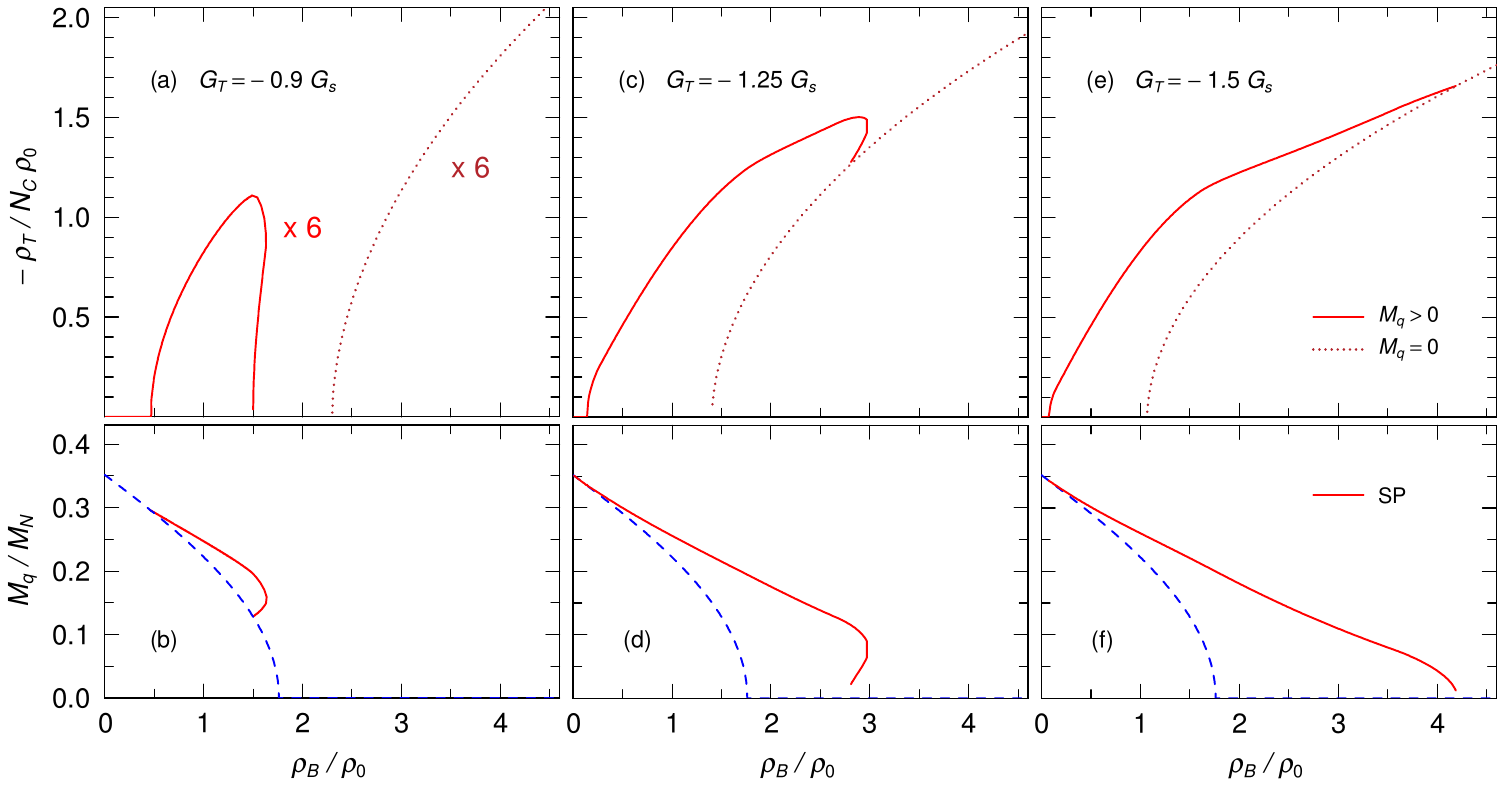}}
\caption{\small 
T-type spin polarization  characteristics with PM2.
In the upper panels (a, c, e), the tensor densities are normalized to the normal nuclear density.
The solid and dotted lines correspond to the results in the massive quark and massless quark phases, respectively.
In the lower panels (b, d, f), the dynamical quark mass is normalized 
by the nucleon mass for the spin-polarized (red solid lines) and spin-saturated (blue dashed line) phases.
The left, middle and right panels show the results for $G_T = -0.9 G_s$ (a, b), $=-1.25 G_s$ (c, d), and $=-1.5 G_s$ (e, f), respectively.
}
\label{TSPM2}
\end{figure}

In Figure~\ref{TSPM2}, we show the results with the parameter set  PM2.
 for $G_T = - 0.9 G_s$ (a,b), $G_T = -1.25 G_s$ (c,d),  and $G_T = -1.5 G_s$ (c,d).
$-G_T / G_s = 0.9$ is close to the lowest coupling strength for the SP,
and the TS-I state  exist in $\rho_B < \rhChN$, and the TS-II state appears
in $\rho_B > \rhChN$, though the two TS-I states exist in some density region, 

As the tensor coupling strength increases, the TS-I state extends into a larger density region,
$G_T = -1.25 G_s$ and $G_T = -1.5 G_s$ the TS-I states extend over 
a density region beyond the  chiral critical density, and the end-points are on the TS-II states,
though there is a density region where the two TS-I exist for  $G_T = -1.25 G_s$.

\bigskip

When $G_A = G_T$, the SP appears at lower density with the T interaction than that with the AV interaction.
The lowest coupling strength for the SP is also smaller in the T interaction than that in the AV interaction. 

\begin{figure}[bht]
\begin{center}
{\includegraphics[scale=0.4,angle=270]{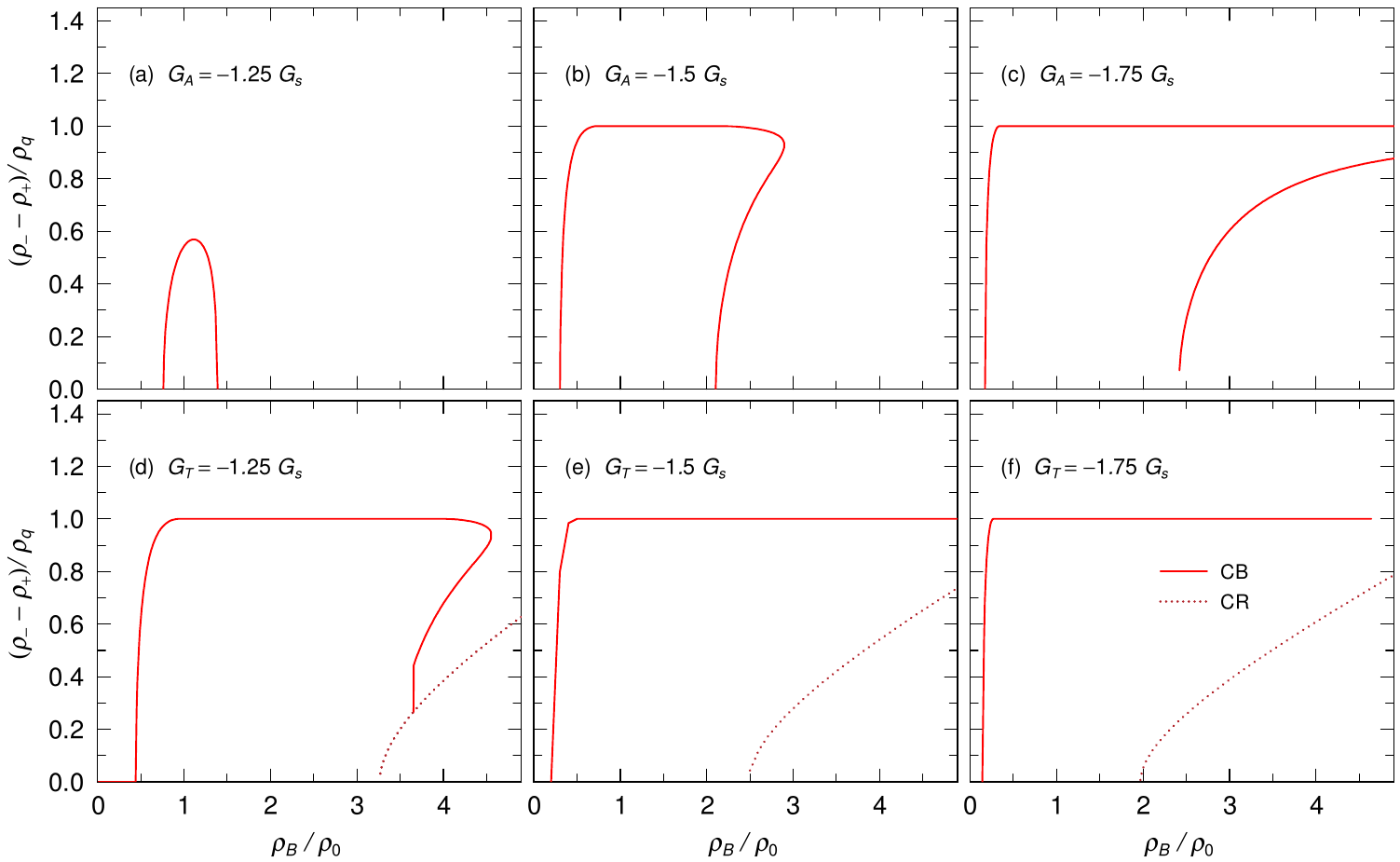}}
\caption{\small 
Spin polarization ratio with PM1.
In the upper panels, the results for the AV type 
when $G_A = -1,25 G_s$ (a),  $G_A = -1,5 G_s$ (b) and  $G_A = -1,75 G_s$ (c).  
In the lower panels   the results for the AV type 
when $G_T = -1,25 G_s$ (c),  $G_T = -1,5 G_s$ (d) and  $G_A = -1,75 G_s$ (e).  
In the  lower panel the solid and dotted lines represent the results 
in the massive and massless spin-polarized phases, respectively.}
\label{SpNP}
\end{center}
\end{figure}

In both types of the SP, when the interactions are weak,  
spin-polarized phases with a finite quark mass appear in the density region below $\rhChN$. 
As the interactions become stronger, a backbending appears in the curve 
representing the spin-polarized states, and two states exist for one density.
When the interaction becomes further stronger
 the spin-polarized phase extends into larger density region,
the two states appear at the same density.
As the interaction becomes further stronger, these two states extend over $\rhChN$.

In the case of the AV type interaction, the massless spin-polarized phase does not exist, and the end density  of 
the spin-polarized phase is in the density region below $\rhChN$.
The curve representing AVS phase shows a backbending even beyond $\rho^{\rm NQ}_c$, 
and the end point is always in the CB density region in normal quark matter ($<\rhChN$).

In the T type interaction, the massless spin-polarized phase does exist,
When the interaction is not very large, the curve representing the state TS-I 
still shows a backbending, but when the interaction is larger, 
it ends at the intersection with the state TS-II.

In this approach we discard the vacuum contribution to the tensor density
though the scalar density includes it,
so that we cannot define the total energy and cannot determine 
what is realized among the spin-saturated (SS), TS-I and TS-II phases.

Next, we examine the occupation number of each spin state in the spin-polarized state.
In Figure~\ref{SpNP} we show the spin polarization ratio $(\rho_{-} - \rho_{+})/\rho_q$
with PM1 for the AV type in the upper panels and for the T type in the lower panels.
As the density increases and exceeds the critical density, 
the spin polarization increases rapidly, reaches one, remains there for a while, 
and then decreases rapidly.
In the TS-II state, on the other hand, the spin polarization increases relatively slowly.

The energy contribution  by the coupling with  the magnetic field $B$  is proportional to 
$\epsilon_{mag} = \mu_q B \rho_T$ with
$\mu_q$ being the quark magnetic moment,
respectively, because   the strength of the magnetic field is  weak in
the energy scale of the strong interaction.
Thus the T density  $\rho_T$ is directly related to the magnetic properties of compact stars,
and we need to examine the T density in the AVS phase.
The T density in the AVS phase is described as
\begin{equation}
\rho_T = N_d \sum_{s=\pm 1} 
 \int \frac{d^3 p}{(2 \pi)^3} n \left[e(\vp,s)  \right] \frac{M_q}{\sqrt{p_z^2 + M_q^2}}. ~~ 
\end{equation}

\begin{figure}[hbt]
\begin{center}
\vspace{-0.5cm}
{\includegraphics[scale=0.32,angle=270]{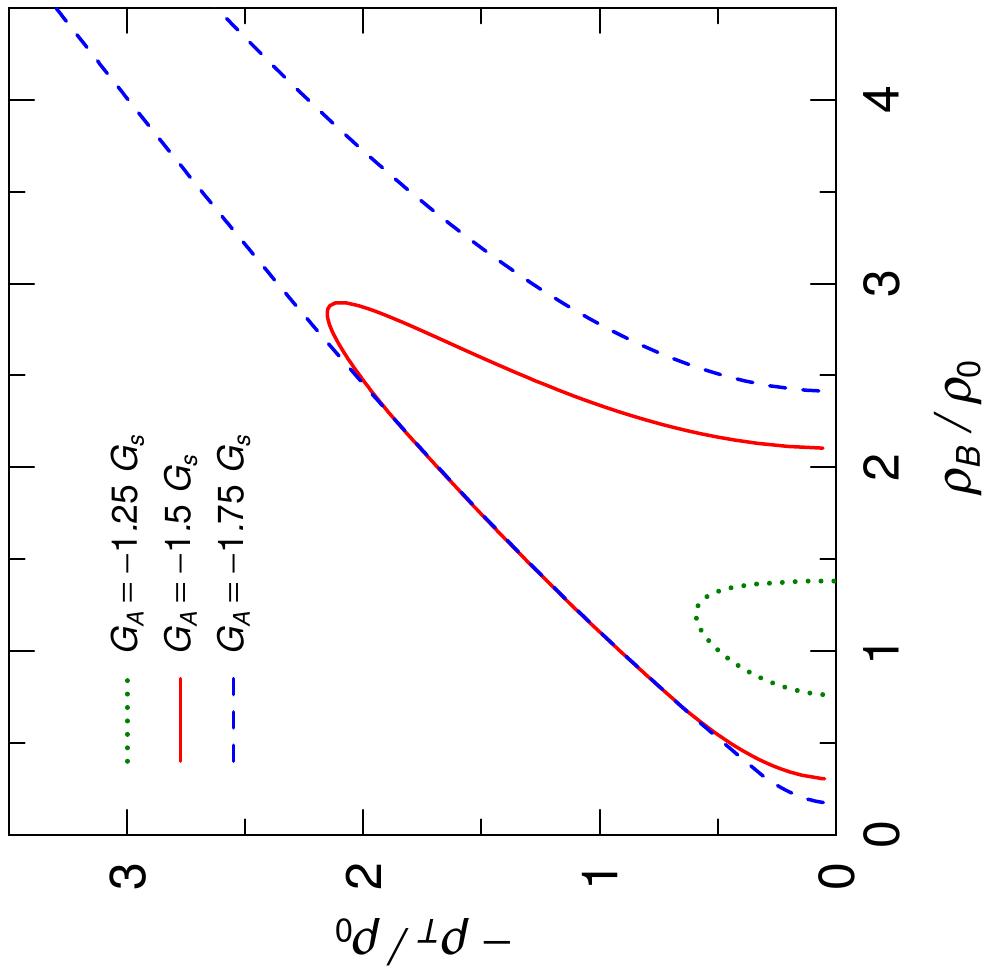}}
\caption{\small Tensor  density in the AV-type spin-polarized phase with 
$G_A / G_s = -1.25$ (dotted line),  $= -1.5$ (solid line) and  $= -1.75$ (dashed line)
in the PM1 prameter-sets.}
\label{TrhoAl}
\end{center}
\end{figure}

In Figure~\ref{TrhoAl} we show the T density in the AV-type spin-polarized system with PM1.
The dotted, solid, dashed lines represent the results for $G_A / G_s = -1.25$,  $G_A / G_s= -1.5$  
and $G_A / G_s = -1.75$, respectively.
We see that the results for $G_A / G_s= -1.5$  and $G_A / G_s= -1.75$ 
agree each other in $0.7 \lesssim \rho_B / \rho_o \lesssim 2.2$,
where the spin polarizations ratio is one for both parameters. 
Furthermore, the T density is larger than the AV density around the peak density region,
and they are also larger than those in the T-type when $G_T = G_A$.
Thus we see that the AV spin-polarization can give rise to  large ferromagnetism.  

The magnetic properties of the quark matter are determined by the two elements:
one is the size of the spin polarization, and the other is the form of the Dirac spinor.
In the variational approach \cite{MaruTatsu00}  the two SP modes, the AV-type and T-type
were demonstrated to be decoupled in the massless limit.
The T density is very small in the AV-type spin-polarized phase, and the AV density is small in
the T-type spin-polarized phase.
By using the exact solutions for the Dirac spinor, we may conclude that both the AV and T densities are large in 
any spin-polarized state.


\bigskip

Finally, we discuss the interplay between spin polarization and chiral symmetry.

In a spin-polarized system, spherical symmetry and inversion symmetry along the $z$-axis are broken, 
but rotational symmetry around the $z$-axis remains. 
The isoscalar tensor interaction breaks chiral symmetry, though this symmetry can be retained 
when it is combined with the isovector tensor interaction.
However, when combined with the isovector tensor interaction, this symmetry can be retained. 
As a result, T-type spin-polarized states exhibit spontaneous chiral symmetry breaking, even when quarks have no mass.

For weak spin-spin interactions—whether AV or T-type—spin-polarized states do not emerge. 
Here, chiral symmetry is spontaneously broken, and quarks acquire a dynamical mass. 
As density increases, this symmetry is restored beyond a critical density, consistent with standard discussions.

Conversely, strong spin-spin interactions lead to spin-polarized states above a critical density, causing chiral symmetry breaking. In this case, a chiral-breaking phase persists beyond $\rho_{\text{ch}}$. For T-type interactions, both massless and massive phases can exist, while for AV-type interactions, two massive phases arise. 
Our findings indicate that strong spin-spin interactions extend the chiral-broken phase to higher densities than in the absence of such interactions.

\section{Summary and Remark}
In this paper we have studied the spontaneous spin-polarized  states caused 
by the axial-vector (AV) mean-field and those caused by the tensor (T) mean field in the NJL model. 
If the AV (T) coupling is beyond  a certain critical strength, 
the AV- (T-)type  spin-polarized phase appears.
As this coupling strength increases, the spin-polarized phase extends into a higher density, 
and as it further increases the spin-polarized phase extends beyond a chiral critical density in normal quark matter..
This spin-polarized phase is a kind of phases where the chiral symmetry is spontaneously broken.

 The AV interaction has chiral symmetry and cannot, on its own, cause a phase 
 with spontaneous chiral symmetry breaking. 
 In the AV-type spin-polarized phase, the dynamical quark mass stays non-zero, so this phase has two massive states above the chiral critical density.
  As the density increases, the spin-polarized phase extends 
  into a higher density range beyond  $\rhChN$,
 
On the other hand, the T interaction consists of two terms — isoscalar and isovector — 
which together maintain chiral symmetry (see Eq.~(\ref{TNcn})). 
However, only one of these terms can produce a mean field that spontaneously breaks chiral symmetry.
In the T-type interaction, not only does the TS-I state extend to high density, but the TS-II state also emerges, as the dynamical quark mass is not required for the formation of the chiral symmetry breaking phase.
 In this approach, the TS-II state exists across the high-density region without a clear boundary.

In the spin-polarized phase, spontaneous breaking of chiral symmetry occurs. 
This phase extends into the high-density region beyond the critical chiral density in normal quark matter, 
regardless of whether the spin-spin interaction is of the axial-vector (AV) or tensor (T) type. 
These findings suggest that spontaneous chiral symmetry breaking, whether due to AV or T-type interactions, 
is realized at higher densities than that in normal quark matter predicted by the standard NJL model.

In this work, multiple spin-polarized states emerge in the results. 
Determining which state is physically realized and whether the phase transition is first-order or second-order requires calculating and comparing the values of the total energy of the quark matter. 
However, this calculation cannot be performed in the current study due to the inability to incorporate the vacuum polarization effects in the AV and T interactions. 
This is because the contribution from vacuum polarization in the AV and T mean fields depends 
on the regularization scheme used.

The significant vacuum effect on spin polarization suggests the presence of large antimagnetic behavior, 
which is inconsistent with lattice QCD results. 
However, even after removing the vacuum effect with $U_T^2$, the counterterm largely grows as $U_T$ 
increases,  preventing us from obtaining reasonable results.

In this work we study  the AV-type and T-type spin-polarized states, but not the states 
with coexistence of both the AV and T mean-fields because its calculation is too hard to be performed 
when the dynamical quark mass is non-zero. 
However, we have calculated this mixed states in the zero quark mass system, and found
the AV mean-field can appear in the density region of the TS-II state \cite{MaruNaka18} .
As mentioned before, the T mean-field makes the spontaneous breaking of the chiral symmetry, 
and the AV field can appear in such a phase. 
Thus,  the T fields as well as the scalar field play the roles to make the spontaneous chiral symmetry breaking,
and the AV field can appear in this field.

In this work, we do not introduce effects of explicitly breaking of the chiral symmetry such as a current massand external magnetic field. These effects are needed for the realistic discussion of the possible 
spin polarization inside compact stars.



\section*{Acknowledgement}
This work is supported by Grants-in-Aid for Scientific Research of JSPS (JP19K03833, JP24K07057)


\bigskip

\appendix

\section{Density Dependent Parts of Densities}
\label{AsecRh}

In this section, we descibe the detailed expressions of the quark number density
$\rho_q$ , the scalar density $\rho_s$, the AV density $\rho_A$ and the T density $\rho_T$ 
with the quark mass $M_q$, the chemical potential $\mu$ and
the AV mean field  $U_A (>0)$ or the T mean ield $U_T (>0)$.

Here, we define an arbitrary operator $\cA (\vp)$ in the momentum space, which has 
the rotation symmetry around the $z$-axis and depends on the $p_z$ and $p_T = |\vp_T|$.
Then, we consider the integration of $F(p_T, p_z)$ which has   
over the momentum inside the Fermi sea, which is described as
\begin{eqnarray}
< \cA > &=& \sum_{s = \pm1 }
\int d^3 \vp ~ \Theta \left[ \mu - e(\vp, s) \right] F( p_z,p_T  ) .
\end{eqnarray}

When $U_A \neq 0$ and $U_T = 0$, the single particle energy is  described as
\begin{equation}
e(\vp_T, p_z ,s) = \sqrt{(\sqrt{m^2+p_z^2} + s U_A )^2 + \vp_T^2} .
\end{equation}
Then,
\begin{equation}
<\cA> = 
 2 \pi  \sum_{s = \pm1 } \int^{\sqrt{(\mu -s U_A)^2 - M_q^2}}_{p_z^{min}} d p_z 
 \int_0^{\sqrt{\mu^2 -(\sqrt{p_z^2 + M_q^2} + s U_A )^2} }  d p_T p_T F(p_z, p_T) ,
\end{equation}
where
$p_z^{min} = \sqrt{ (U_A - \mu)^2 - M_q^2}$ when $\mu <  M_q +  U_A$ and $s=-1$, and
$p_z^{min}=0$ otherwise.

When $U_T \neq 0$ and $U_A = 0$, the single particle energy is described as
\begin{equation}
e(\vp_T, p_z ,s) = \sqrt{(\sqrt{m^2+p_T^2} + s U_T )^2 + \vp_z^2} .
\end{equation}
Then,
\begin{equation}
<\cA> = 
 2 \pi  \sum_{s = \pm1 } \int^{\sqrt{(\mu -s U_A)^2 - M_q^2}}_{p_T^{min}} d p_T p_T 
 \int_0^{\sqrt{\mu^2 -(\sqrt{p_T^2 + M_q^2} + s U_T )^2} }  d p_z  F(p_z, p_T) ,
\end{equation}
where
$p_T^{min} = \sqrt{ (U_T - \mu)^2 - M_q^2}$ when $\mu <  M_q +  U_T$ and $s=-1$, and
$p_T^{min}=0$ otherwise.


\subsection{Axial-Vector Type Spin Polarization}
\label{SecDDAV}

In this subsection we show the detailed expressions of the densities when $U_A \ge 0$ and $U_T = 0$.

\subsubsection{Quark Number Density}

The quark number density, $\rho_q$ is written by 
\begin{equation}
\rho_q (s) = \frac{N_d}{(2 \pi)^3} \int d^3 p  ~ n \left(  e (\vp,s) \right) .
\end{equation}
with $n(E) = \Theta \left[ \mu - E\right] $

When $\mu \ge M_q + sU_A$, , 
\begin{eqnarray}
\rho_q (s) &=&  
%
  \frac{ N_d}{(2 \pi)^2} 
\left\{   \frac{\sqrt{(\mu -s U_A)^2-M_q^2}}{3}  
\right. \nonumber \\ && \left. \qquad\qquad 
\times
 \left[ 2 ( \mu - s U_A ) (  \mu  + 2 s U_A ) - 3 s U_A \left| \mu - sU_A \right| - 2 M_q^2 \right]
\right. \nonumber \\ &&  \left. \qquad   
- \frac{s}{2} U_A M_q^2 
\ln \left( \frac{ \mu -s U_A + \sqrt{(\mu -s U_A)^2 - M_q^2}  }{\mu -s U_A - \sqrt{(\mu -s U_A)^2 - M_q^2} } \right)
 \right\}  \equiv \rho_q^{(0)} .
\end{eqnarray}

%
%

When $U_A +s M_q \ge \mu$, 
\begin{eqnarray}
\rho_q (+) &=& 0,
%
\\
\rho_q (-) &=&  \rho_q^{(0)}
+ \frac{ N_d}{(2 \pi)^2} \left\{  \frac{ 2 \sqrt{(U_A - \mu )^2-M_q^2 } }{3} (U_A - \mu) (2 \mu + U_A ) 
 \right. \nonumber \\ && \left.. \qquad\qquad\qquad
 - \frac{1}{2}M_q^2  U_A 
    \ln \left[ \frac{ U_A - \mu + \sqrt{(U_A - \mu )^2 - M_q^2 } }{ U_A - \mu - \sqrt{(U_A - \mu )^2 - M_q^2 } } \right] 
\right\} .
\end{eqnarray}
%

\subsubsection{Scalar Density}
The scalar density is described as
\begin{eqnarray}
 \rho_s &=& 
N_d \sum_{s=\pm 1} \int \frac{d^3 p}{(2 \pi)^3} 
 n(e(\vp,s))\frac{ M_q \sqrt{ p_z^2 + M_q^2 } + s M_q U_A }{ e \sqrt{ p_z^2 + M_q^2 } } 
\nonumber \\ &=& 
 \frac{ M_q }{ (2 \pi)^2} \sum_{s=\pm 1} \int d p_z \left[ 1 + \frac{ s U_A }{\sqrt{M_q^2+p_z^2} }  \right]
\nonumber \\ && \qquad\qquad\qquad  
\times
\int_0^{ \sqrt{\mu^2 - (E_z + s U_A)^2} } \frac{ d p_T p_T }{\sqrt{p_T^2 + ( E_z + s U_A )^2 } },
\end{eqnarray}
with $E_z=\sqrt{ p_z^2 + M_q^2 }$.
When $M_q \ge U_A$ and $s-1$, or $\mu \ge M_q + U_A$ and $s=+1$,
\begin{eqnarray}
 \rho_s (s) &=&  
\frac{N_d M_q }{(2 \pi)^2}  
\left\{ \frac{1}{2}  (\mu  - 3 s U_A) \sqrt{(\mu -s U_A)^2-M_q^2} 
\right. \nonumber \\ && \quad \left.
+ \left( s U_A \mu - U_A^2 - \frac{M_q^2}{2} \right) 
 \ln \left[ \frac{ (\mu -s U_A) + \sqrt{ (\mu -s U_A)^2  - M_q^2 }  }{(\mu -s U_A) - \sqrt{ (\mu -s U_A)^2  - M_q^2 } } \right] 
\right\} , \qquad
\end{eqnarray}
%

When $M_q \le U_A \le \mu + M_q$ and $s=-1$, 
\begin{eqnarray}
 \rho_s (-1)  &=&  
\frac{N_d M_q }{(2 \pi)^2}  
\left\{  (\mu + 3 U_A ) \sqrt{(\mu + U_A)^2-M_q^2} - 6 U_A \sqrt{U_A^2 - M_q^2}  
\right. \nonumber \\ && \quad \left.
+ \left( M_q^2 + 2 U_A^2 \right)  \ln \left[ \frac{ U_A  + \sqrt{U_A^2 - M_q^2} }{ U_A  - \sqrt{U_A^2 - M_q^2} } \right] 
%
\right. \nonumber \\ && \quad \left.
-  \left[ \frac{M_q^2}{2}  + U_A (\mu + U_A )  \right]  
     \ln \left[ \frac{ \mu + U_A  + \sqrt{(\mu + U_A)^2-M_q^2} }{ \mu + U_A  - \sqrt{(\mu + U_A)^2-M_q^2  }} \right]
 \right\} . \qquad
\end{eqnarray}


When $U_A +s M_q \ge \mu$, 
%
%
\begin{eqnarray}
 \rho_s (+1)  &=& 0,
\\ 
 \rho_s (-1)  &=&
\frac{N_d M_q}{(2 \pi)^2}  
\left\{ (\mu + 3 U_A ) \sqrt{(\mu + U_A)^2-M_q^2}  +  ( 3 U_A - \mu ) \sqrt{(U_A - \mu )^2-M_q^2}
\right. \nonumber \\ && \quad 
+  ( 3 U_A - \mu ) \sqrt{(U_A - \mu )^2-M_q^2}
%
 \nonumber \\ && \quad 
 - 3 U_A \sqrt{U_A^2 - M_q^2} + M_q^2 \ln\left(\frac{ U_A  + \sqrt{U_A^2 - M_q^2} }{ U_A  - \sqrt{U_A^2 - M_q^2} } \right) 
\nonumber \\ && \quad 
-  \left( M_q^2 + 2 U_A^2 + 2 U_A \mu \right)    
\ln \left( \frac{ \sqrt{(  U_A + \mu )^2-M_q^2} + U_A + \mu  }{M_q} \right)
 \nonumber  \\ && \quad \left.
 -  \left( M_q^2 + 2 U_A^2 - 2 U_A \mu \right)    
 \ln \left( \frac{ \sqrt{(  U_A - \mu )^2-M_q^2} + U_A - \mu }{M_q} \right) 
 \right\} .
\end{eqnarray}

\subsubsection{Axial-Vector density}
The AV density is described as
\begin{eqnarray}
 \rho_A (s)
 &=&  N_d \int \frac{d^3 p}{(2 \pi)^3} 
 n( e(\vp,s) ) \frac{ s \sqrt{p_z^2 + M_q^2} + U_A }{ e(\vp,s)} .
\label{AvRhF}
\end{eqnarray}

When $M_q \ge U_A$ and $s-1$ or $\mu \ge M_q + U_A$ and $s=+1$,
\begin{eqnarray}
 \rho_A (s)  &=&
 \frac{N_d }{ (2 \pi)^2}   s 
\left\{  \frac{ 1 }{3} \sqrt{ (\mu -s U_A)^2 - M_q^2 }\left[ (\mu + 2sU_A) (\mu - s U_A) - 4 M_q^2 \right]
\right. \nonumber \\ && \left. \qquad
+   \frac{M_q^2}{2} ( \mu -  2s U_A  ) \ln \left(\frac{ \mu - s U_A + \sqrt{(\mu -s U_A)^2-M_q^2} }{\mu - s U_A - \sqrt{(\mu -s U_A)^2-M_q^2} } \right)
\right\} .
\end{eqnarray}

When $M_q \le U_A \le \mu + M_q$ and $s=-1$,
\begin{eqnarray}
 \rho_A (-) &=&
- \frac{ N_d }{ (2 \pi)^2}   
\left\{ \frac{ \sqrt{ (\mu + U_A)^2 - M_q^2 } }{3} \left[ (\mu - 2 U_A) (\mu + U_A) - 4 M_q^2 \right]
\right. \nonumber \\ &&  \qquad
+ \frac{ \sqrt{ U_A^2-M_q^2} }{ 3 } \left(  U_A^2  + 2 M_q^2 \right) 
 - 2 M_q^2 U_A \ln \left(\frac{ U_A + \sqrt{U_A^2-M_q^2} }{ U_A - \sqrt{U_A^2-M_q^2} } \right) 
%
\nonumber \\ && \left. \qquad
 +  \frac{M_q^2}{2} ( \mu + 2 U_A  ) \ln \left(\frac{ \mu + U_A + \sqrt{(\mu + U_A)^2-M_q^2} }{ \mu + U_A - \sqrt{(\mu + U_A)^2-M_q^2}  } \right)
\right\}  . \qquad
\end{eqnarray}
%
  
When $U_A +s M_q \ge \mu$, 
\begin{eqnarray}
 \rho_A (+)  &=&  0,
 %
 %
\\ \rho_A (-)  &=&
- \frac{ 2 N_d }{ (2 \pi)^2}   
\left\{ \frac{ \sqrt{ (\mu + U_A)^2 - M_q^2 } }{6} \left[(\mu + U_A)  (\mu - 2 U_A) - 4 M_q^2 \right]
\right. \nonumber \\ && \qquad\quad~~
+ \frac{2 }{3} \sqrt{ U_A^2-M_q^2} \left(  U_A^2  + 2 M_q^2 \right) 
 \nonumber \\ && \qquad\quad~~
 + \frac{\sqrt{(U_A - \mu )^2 -M_q^2}}{6} \left[ (\mu - U_A  ) ( \mu + 2 U_A )  - 4 M_q^2 \right]
\nonumber  \\ && \qquad\quad~~
 - 2 M_q^2 U_A \ln \left(\frac{U_A + \sqrt{U_A^2-M_q^2}}{m} \right) 
\nonumber  \\ && \qquad\quad~~
 +  \frac{M_q^2}{2} ( \mu + 2 U_A  ) \ln \left(\frac{ \mu + U_A + \sqrt{(\mu + U_A)^2-M_q^2} }{M_q} \right)
\nonumber  \\ && \left.  \qquad\quad~~
- \frac{M_q^2}{2}  ( \mu -  2 U_A  ) \ln \left(\frac{ U_A - \mu + \sqrt{ (U_A - \mu )^2-M_q^2} }{M_q} \right) 
\right\} .
\end{eqnarray}
%
%

\subsection{Tensor Type Spin Polarization}
\label{SecDDTS}

\subsubsection{Quark Number Density}

When $U_T < \mu - s M_q$ for $s=\pm 1$,
\begin{eqnarray}
\rho_q (s)  &=&  \frac{N_d}{2 \pi^2} \left\{ 
\frac{1}{6} \sqrt{\mu^2 - (M_q+sU_T)^2 }
\left[ 2 \mu^2- (M_q+sU_T)(2M_q-sU_T) \right] 
\right. \nonumber \\ && \left. \quad \quad 
- \frac{s}{2} U_T \mu^2 \left[ \frac{\pi}{2}
 - \sin^{-1} \left( \frac{M_q+sU_T}{\mu} \right) \right] \right\} ,
\label{qRh-A2}
\end{eqnarray}

When $U_T > \mu - s M_q$,
\begin{equation}
\rho_q (+1) = 0 , \quad 
\rho_q (-1)  =    \frac{N_d}{4 \pi} U_T  \mu^2 .
\label{qRh-A2b}
\end{equation}

\bigskip

\subsubsection{Scalar Density}

When $U_T < \mu - s M_q$ for $s=\pm 1$,
\begin{eqnarray}
\rho_s (s) &=& 
\frac{N_d}{4 \pi^2} M_q  \left[ \mu \sqrt{\mu^2 -(M_q+sU_T)^2} \right.
\nonumber \\ && \quad   \quad   \quad  
\left. - \frac{(M_q+sU_T)^2}{2} \ln \left( 
\frac{\mu + \sqrt{\mu^2 -(M_q+sU_T)^2}}{\mu - \sqrt{\mu^2 -(M_q+sU_T)^2}} \right)
\right] ,
\label{RhS-A2}
\end{eqnarray}


When $U_T > \mu - s M_q$,
\begin{equation}
 \rho_s (\pm 1) = 0 ,
\end{equation}

\bigskip

\subsubsection{Tensor Density}

When $U_T < \mu - M_q$ for $s=1$ or $U_T < M_q$ for $s=-1$ 
\begin{eqnarray}
\rho_T (s) &=& \frac{N_d}{12 \pi^2} s \left\{ 
(M_q+sU_T)^2 \left( -  M_q + \frac{s}{2}U_T \right)
\ln \left( \frac{\mu + \sqrt{\mu^2 -(M_q+sU_T)^2}}
{\mu - \sqrt{\mu^2 -(M_q+sU_T)^2}} \right)
\right. \nonumber \\ && \left. \quad\quad\quad\quad
+ ~ \mu (M_q-2sU_T) \sqrt{\mu^2 -(M_q+sU_T)^2}
\right. \nonumber \\ && \left. \quad\quad\quad\quad
+ \mu^3 \sin^{-1} \left(\frac{\sqrt{\mu^2 -(U_T + sM_q)^2}}{\mu}\right) \right\} .
\label{TRh-A2}
\end{eqnarray}
When $U_T > \mu -M_q$ for $s=1$,
\begin{equation}
\rho_T(+1) = 0 .
\end{equation}

When $\mu + M_q > U_T > M_q$ for $s=-1$,
\begin{eqnarray}
\rho_T (-1) &=& - \frac{N_d}{12 \pi^2} \left\{
- \frac{1}{2} (U_T - M_q)^2 \left( U_T + 2M_q \right)
\ln \left( \frac{\mu + \sqrt{\mu^2 -(U_T - M_q)^2}}
{\mu - \sqrt{\mu^2 -(U_T - M_q)^2}} \right)
\right. \nonumber \\ &&  \quad\quad\quad\quad
+ ~ \mu \left( 2U_T + M_q \right) \sqrt{\mu^2 -(U_T - M_q)^2}
\nonumber \\ && \left .\quad\quad\quad\quad
+ ~ \mu^3 \left[ \pi -  \sin^{-1} 
\left(\frac{\sqrt{\mu^2 -(U_T-M_q)^2}}{\mu} \right) \right]  \right\} .
\label{TRh-A2a}
\end{eqnarray}
When $U_T > \mu + M_q$  for $s=-1$, 
\begin{eqnarray}
\rho_T (-1) &=&  - \frac{N_d}{12 \pi}  \mu^3 .
\label{TRh-A2b}
\end{eqnarray}

\newpage

\section{Vacuum Contributions of Densities}
\label{BsecRh}

\subsection{Axial-Vector Type Spin Polarization}
\label{SecVCAV}

When $U_A >0$ and $U_T = 0$,
the termodynamical potential is written by
\begin{eqnarray}
\Omega_{vac} & = &   i N_d \sum_{s = \pm 1} \int \frac{d^4 p}{(2 \pi)^4} \int_{1/\Lambda^2}^{\infty} 
\frac{d \tau}{\tau} e^{\tau \left[ p_0^2 - p_T^2 - ( \sqrt{p_z^2 + M_q^2} + s U_A)^2.\right] }
%
%
\nonumber \\  &=&
 \frac{1}{ 8 \pi^{5/2} } \sum_{s = \pm 1} \int_{1/\Lambda^2}^{\infty} \frac{d \tau}{\tau^{5/2}}  
\int_{0}^{\infty} d p_z e^{- \tau \left( \sqrt{p_z^2 + M_q^2} + sU_A \right)^2 } .
\end{eqnarray}

Then, the scalar density is expressed as
\begin{eqnarray}
&& \rho_s(V) = \frac{ \partial \Omega_{vac}}{\partial M_q }
%
\nonumber \\  &=&  
 \sum_{s = \pm 1} \frac{-2 N_d}{ 8 \pi^{5/2} }  \int_{1/\Lambda^2}^{\infty} \frac{d \tau}{\tau^{3/2}}  
 \int_{0}^{\infty} d p_z  \frac{M_q}{\sqrt{p_z^2 + M_q^2} } \left( \sqrt{p_z^2 + M_q^2} + sU_A \right)
\nonumber \\  && \qquad\qquad\qquad\qquad\qquad\qquad\qquad\qquad \times  
  e^{- \tau \left( \sqrt{p_z^2 + M_q^2} + sU_A \right)^2 } .
%
\nonumber \\  &=&  
- \frac{N_d \Lambda }{ 4 \pi^{5/2} } M_q \sum_{s = \pm 1}  \int_{0}^{\infty} d p_z    
 \left( 1 + \frac{sU_A}{\sqrt{p_z^2 + M_q^2} } \right)
 F_{3/2} \left[ \left| \frac{ \sqrt{p_z^2 + M_q^2} + sU_A }{\Lambda} \right| \right] . \qquad 
 \end{eqnarray}
The AV density is expressed as
\begin{eqnarray}
&& \rho_A(V) = \frac{ \partial \Omega_{vac}}{\partial U_A}
%
\nonumber \\  &=&  
\sum_{s = \pm 1} \frac{-2s N_d}{ 8 \pi^{5/2} }  \int_{1/\Lambda^2}^{\infty} \frac{d \tau}{\tau^{3/2}}  
 \int_{0}^{\infty} d p_z  \left( \sqrt{p_z^2 + M_q^2} + sU_A \right)
  e^{- \tau \left( \sqrt{p_z^2 + M_q^2} + sU_A \right)^2 } 
%
\nonumber \\  &=&  
  \frac{- N_d \Lambda}{ 4 \pi^{5/2} }  \sum_{s = \pm 1}  s
 \int_{0}^{\infty} d p_z  \left( \sqrt{p_z^2 + M_q^2} + sU_A \right)
 F_{3/2}  \left[ \left| \frac{ \sqrt{p_z^2 + M_q^2} + sU_A }{\Lambda} \right| \right] , \qquad
\end{eqnarray}
where $F_\alpha (x)$ is defined in Eq.~(\ref{Ffnc}).
%
%

\subsection{Tensor Type Spin Polarization}
\label{SecVCTS}

The vacuum contribution to the thermodynamic potential is described as 
\begin{eqnarray}
\Omega_{vac} & = &   i N_d \sum_{s = \pm 1} \int \frac{d^4 p}{(2 \pi)^4} \int_{1/\Lambda^2}^{\infty} 
\frac{d \tau}{\tau} e^{\tau \left[ p_0^2 - p_z^2 - ( \sqrt{p_T^2 + M_q^2} + s U_T)^2.\right] }
%
\nonumber \\  &=&
 \frac{N_d}{8 \pi^2} \sum_{s = \pm 1} \int_{1/\Lambda^2}^{\infty} 
\frac{d \tau}{\tau^2} \int_{0}^{\infty} d p_T p_T 
e^{- \tau (\sqrt{p_T^2 + M_q^2} + s U_T)^2 } 
%
\nonumber \\  &=&
 \frac{N_d}{8 \pi^2} \sum_{s = \pm 1} \int_{1/\Lambda^2}^{\infty} 
\frac{d \tau}{\tau^2} \int_{M_q}^{\infty} d E_T E_T 
e^{- \tau (E_T + s U_T)^2 } .
\end{eqnarray}
Then, the  scalar density is expressed as 
\begin{eqnarray}
&& \rho_s (V)  =  \frac{ \partial \Omega_{vac} }{ \partial M_q~~} 
\nonumber \\ &=&
 - \frac{N_d}{4 \pi^2} M_q \sum_{s = \pm 1} \int_{1/\Lambda^2}^{\infty} 
\frac{d \tau}{\tau} \int_{M_q}^{\infty} d E_T \left(  E_T + s U_T \right) 
e^{- \tau ( E_T + s U_T)^2 } .
\nonumber \\ &=&
 - \frac{N_d M_q}{8 \pi^2} \sum_s \int_{1/\Lambda^2}^{\infty} 
\frac{d \tau}{\tau^2} e^{- \tau (M_q + s U_T)^2 } 
=  - \frac{N_d M_q}{8 \pi^2} \Lambda^2 \sum_s
F_2 \left( \frac{(M_q + s U_T)^2}{\Lambda^2} \right) . \qquad
\end{eqnarray}
The tensor density is expressed as
\begin{eqnarray}
&& \rho_T (V)  = \frac{ \partial \Omega_{vac} }{ \partial U_T~~} 
%
\nonumber \\ &=&
- \frac{N_d}{4 \pi^2} \sum_{s = \pm 1} s \int_{1/\Lambda^2}^{\infty} 
\frac{d \tau}{\tau} \int_{M_q}^{\infty} d E_T E_T (E_T + s U_T)
e^{- \tau (E_T + s U_T)^2 } 
%
\nonumber \\ &=&
- \frac{N_d}{4 \pi^2} \sum_{s = \pm 1} s \int_{1/\Lambda^2}^{\infty} 
\frac{d \tau}{\tau} \int_{M_q + s U_T}^{\infty} d E_T E_T (E_T - s U_T) e^{- \tau E_T^2 } 
%
%
\nonumber \\ &=&
- \frac{N_d}{4 \pi^2} \sum_{s = \pm 1} s \int_{1/\Lambda^2}^{\infty} \frac{d \tau}{\tau} 
\left\{ 
\int_{M_q + s U_T}^{\infty} d E_T E_T^2e^{- \tau E_T^2 } 
- \frac{U_T}{2\tau} e^{- \tau (M_q + s U_T)^2 }     
\right\}
%
\nonumber \\ &=&
\frac{N_d}{8 \pi^2} \Lambda^2 \left\{ 
2 \int_{M_q - U_T}^{M_q + U_T} d E_T F_1 \left( \frac{E_T^2}{\Lambda^2} \right) 
+  \sum_s U_TF_2\left[\frac{(M_q + sU_T)^2}{\Lambda^2}\right] 
\right\} . \qquad\qquad
\end{eqnarray}
%

\vspace{6pt} 



\end{document}